\documentclass[pre,amsmath,amsfonts,nofootinbib,floatfix,twocolumn,showpacs]{revtex4}
 
\newcommand{\PRE}[1]{#1}
\newcommand{\PREdel}[1]{}

\usepackage{graphicx}

\usepackage{subfigure}

\graphicspath{{Figures/}}
 
\newcommand{\figurewidth}{\columnwidth}
\newcommand{\subfigurewidth}{0.5\textwidth}

\usepackage[colorlinks,bookmarks]{hyperref}

\usepackage{dcolumn}
\newcolumntype{=}{D{=}{{}={}}{-1}}
\newcolumntype{~}{D{~}{{}\sim{}}{-1}}
\newcolumntype{,}{D{,}{{}={}}{-1}}
\newcolumntype{|}{D{|}{{}\quad{}}{-1}}
\newcolumntype{x}{D{x}{{}\times{}}{-1}}
 
\usepackage[mediumspace,squaren,textstyle]{SIunits}

\addunit{\Equivalent}{equiv}

\begin{document} 

\title{Phase field modeling of electrochemistry II: Kinetics}

\author{J. E. Guyer}
\email{guyer@nist.gov} 
\author{W. J. Boettinger}
\email{william.boettinger@nist.gov} 
\author{J. A. Warren}
\email{jwarren@nist.gov} 
\affiliation{Metallurgy Division, Materials Science and
Engineering Laboratory, National Institute of Standards and Technology,
Gaithersburg, MD 20899}

\author{G. B. McFadden} 
\email{mcfadden@nist.gov}
\affiliation{Mathematical and Computational Sciences Division, Information
Technology Laboratory, National Institute of Standards and Technology,
Gaithersburg, MD 20899} 
 
\begin{abstract} 
    The kinetic behavior of a phase field model of electrochemistry is explored
    for advancing (electrodeposition) and receding (electrodissolution)
    conditions in one dimension.  We described the equilibrium behavior of this
    model in [J. E. Guyer, W. J. Boettinger, J.A. Warren, and G. B. McFadden,
    ``Phase field modeling of electrochemistry I: Equilibrium'',
    \href{http://arxiv.org/abs/cond-mat/0308173}{cond-mat/0308173}].  We examine the relationship
    between the parameters of the phase field method and the more typical
    parameters of electrochemistry.  We demonstrate ohmic conduction in the
    electrode and ionic conduction in the electrolyte.  We find that, despite
    making simple, linear dynamic postulates, we obtain the nonlinear
    relationship between current and overpotential predicted by the classical
    ``Butler-Volmer'' equation and observed in electrochemical experiments.  The
    charge distribution in the interfacial double layer changes with the passage
    of current and, at sufficiently high currents, we find that the diffusion
    limited deposition of a more noble cation leads to alloy deposition with less
    noble species.
\end{abstract} 


\date{\today} 

\maketitle 

\section{Introduction} 

In Ref.~\cite{ElPhF:equilibrium}, we developed an equilibrium phase field model
of an electrochemical system.  In this paper, we examine the dynamic aspects of
that model.  Models of phase transformations can be broadly categorized into
sharp or diffuse interface approaches.  Sharp interface models treat the
transition between phases as mathematically abrupt.  Diffuse interface models
assume that the phase interface has some finite thickness over which material
properties vary smoothly.  Both cases are simplifications of the physical
interface between phases, in which properties vary over some finite, atomic-scale
distance which is often smaller than assumed in diffuse interface models.
Traditional equilibrium models of electrochemical interfaces take the interface
between phases (the transition between ``electrode'' and ``electrolyte'') to be
abrupt, but frequently consider the distribution of charge and electrostatic
potential to be diffuse in the electrolyte, as by the Gouy-Chapman-Stern model
\cite{Grahame:1947}.  Dynamic models of electrochemistry typically ignore the
details of the charge distribution at the interface and employ a fully sharp
model that we summarize in Section~\ref{sec:Sharp}.

The phase field technique is one particular diffuse interface approach.  The
method employs a phase field variable, which is a function of position and time,
to describe whether the material is one phase or another, \emph{e.g.}, solid or
liquid.  The behavior of this variable is governed by a partial differential
equation (PDE) that is coupled to the relevant transport equations for the
material.  The interface between the phases is described by smooth but highly
localized changes of this variable.  This approach avoids the mathematically
difficult problem of applying boundary conditions at an interface whose location
is part of the unknown solution.  The phase field method is powerful because it
can easily treat complex interface shapes and topology changes.  Our long term
goal is to treat the complex geometry, including void formation, that occurs
during electrodeposition in vias and trenches for on-chip metallization and in
the dendritic structures that form during battery recharging.  Phase field
methods will allow the rigorous examination of the interplay between current,
potential gradients, curvature, and adsorption in intricate geometries.

There is a rich body of literature of sharp interface models of
electrodeposition, which we will sketch briefly in Section~\ref{sec:Sharp}, but
the application of diffuse interface techniques to the motion of electrochemical
interfaces has been relatively limited.  Dussault and Powell have applied phase
field techniques to the modeling of electrochemical processes in steel slags
\cite{Dussault:2002,Powell:2003}, but their approach neglects the effects of
charge at the interfacial double layer.  As a result, they are able to model much
larger domains and much longer time spans than we present here, but the essential
physics of the electrocapillary interface are not examined.  Wheeler, Josell, and
Moffat have performed a level set analysis of so-called ``superconformal''
electrodeposition in high aspect ratio features, with particular emphasis on the
role of additives \cite{NIST:damascene:2003}.  Like phase field models, level set
techniques allow the treatment of complex morphologies, such as the formation of
voids during trench filling, but the motion of the interface is handled
phenomenologically rather than physically, as by the phase field approach, so
again the structure of the double layer is not considered.  Bernard, Plapp, and
Gouyet have recently presented a lattice-gas model of an electrochemical system
\cite{Bernard:2001,Bernard:2003} that exhibits many of the same interfacial and
dynamic behaviors that we find in this paper, as well as exhibiting the early
stages of dendritic growth.  This type of discrete modeling should provide a
useful bridge between an atomistic view of the electrochemical interface and the
continuum approach of phase field models.

To place our results in context, Section~\ref{sec:Sharp} outlines the traditional
sharp interface description of electrodeposition.  Section~\ref{sec:Model} presents
the dynamic postulates governing the evolution of the phase, concentration, and
electrostatic potential fields which we proposed in
Ref.~\cite{ElPhF:equilibrium}.  Section~\ref{sec:Numerics} describes our
numerical approach and boundary conditions.  Section~\ref{sec:Parameters}
discusses the selection of materials parameters, including the relationship
between the phase field mobility and the Butler-Volmer exchange current of
traditional electrochemical modeling.  Section~\ref{sec:Solutions} presents the
results of numerical calculations in one spatial dimension that span a range of
electrodeposition and electrodissolution conditions.

\section{Sharp Interface Approach for Electrodeposition}
\label{sec:Sharp}

In Ref.~\cite{ElPhF:equilibrium}, we present a phase field model of the
equilibrium between an electrode phase \Electrode\ and an electrolyte phase
\Electrolyte, consisting of a set of four charged components, \Electron,
\Cation{+\NumberCa}, \Otherion{+\NumberOt}, and \Anion{-\NumberAn}.  A
superscript \Electrode\ denotes that the quantity is evaluated in the bulk
electrode (metal) phase and a superscript \Electrolyte\ denotes that the quantity
is evaluated in the bulk electrolyte phase.  At equilibrium, the difference in
potential \Potential\ between the electrode and the electrolyte in an 
\Components-component system is given by
\begin{equation}
    \Galvani{}
    = \Potential^{\Electrode} - \Potential^{\Electrolyte}
    = -\frac{\Delta\Chemical{j}{\Standard}}{\Valence{j}\Faraday}
    + \frac{\Gas\Temperature}{\Valence{j}\Faraday}\ln
	\frac{\Fraction{j}^{\Electrolyte}}
	    {\Fraction{j}^{\Electrode}},
	    \qquad j = 1\ldots\Components
    \label{eq:Equilibrium}
\end{equation}
where \( \Delta\Chemical{j}{\Standard} =
\Chemical{j}{\Standard\Electrode}-\Chemical{j}{\Standard\Electrolyte} \) and
\Chemical{j}{\Standard} is the chemical potential of pure component \( j \) in
the respective phase, \Valence{j} is the valence of component \( j \),
\Fraction{j} is the mole fraction of component \( j \), \Faraday\ is Faraday's
constant, \Gas\ is the molar gas constant, and \Temperature\ is temperature.
Eq.~\eqref{eq:Equilibrium} is the generalization, for all of the components, of
the Nernst equation of traditional electrochemical analysis.  The equation is
normally only written for the electroactive species.  In
Ref.~\cite{ElPhF:equilibrium} we explain the origin of an equation for each
component in the system and the relationship between the term proportional to \(
\Delta\Chemical{j}{\Standard} \) and the standard cell potential.

When current is passed through the interface, the potential difference \Galvani{}
shifts.  Alternatively, when a potential difference other than the equilibrium
value is imposed, current will pass and the interface will move.  The shift in
the potential difference across the interface (excluding the ohmic drops across
the bulk phases) is referred to as the overpotential \Overpotential\
\cite{Vetter:1967,Bard:2nd}. 

For an electrochemical system with only one monovalent electroactive species
\Cation{+}, chemical reaction rate theory gives the relationship between current
density \Current\ and total overpotential \Overpotential\ for a planar interface
as \cite{Vetter:1967}
\begin{equation}
    \Current 
    = \Current_{0}
    \left\{
	\exp
	\left[
	    \frac{
		\left(
		    1-\Transfer
		\right)
		\Faraday\Overpotential}{\Gas\Temperature}
	\right]
	-\frac{\Concentration{\Cation{+}}}{\Concentration{\Cation{+}}^{\Thickness}}
	\exp
	\left[
	    -\frac{\Transfer\Faraday\Overpotential}{\Gas\Temperature}
	\right]
    \right\}.
    \label{eq:CurrentOverpotential:General}
\end{equation}
The first term in the curly brackets represents the anodic/oxidizing reaction and
the second term represents the cathodic/reducing reaction.
\Concentration{\Cation{+}} is the concentration of cations \Cation{+} in the
electrolyte at electrode-electrolyte interface and \(
\Concentration{\Cation{+}}^{\Thickness} \) is the bulk electrolyte concentration
of cations \Cation{+} at the edge of the diffusion boundary layer.  The exchange
current density \( \Current_{0} \) and the transfer coefficient \Transfer\
characterize the facility and symmetry of the forward and reverse reactions.  The
current density \( \Current \equiv \vec{\Current} \cdot \vec{\Normal} \), where
the normal \( \vec{\Normal} \) points from \Electrode\ into \Electrolyte\ and \(
\vec{\Current} \) is the current density vector.  Thus, positive values of
\Current\ result in dissolution.

If the diffusion field can be assumed to be linear, the implicit dependence of
\Concentration{\Cation{+}} on \Current\ can be eliminated in
Eq.~\eqref{eq:CurrentOverpotential:General}, giving  
\begin{equation}
    \Current 
    = \Current_{0}
    \left\{
	\exp
	\left[
	    \frac{
		\left(
		    1-\Transfer
		\right)
		\Faraday\Overpotential}{\Gas\Temperature}
	\right]
        -\left(
	    1 - \frac{\Current}{\Current_{\text{lim}}}
        \right)
	\exp
	\left[
	    -\frac{\Transfer\Faraday\Overpotential}{\Gas\Temperature}
	\right]
    \right\}.
    \label{eq:CurrentOverpotential}
\end{equation}
This expression can be rearranged to give \Current\ as an explicit function of
\Overpotential, which is useful for comparison to our phase field results.
Linearity of the concentration profile is appropriate only if the interface
velocity is much less than \( \Diffusivity{\Cation{+}} /
\Thickness_{\Diffusivity{}} \), where \Diffusivity{\Cation{+}} is the diffusivity
of \Cation{+} and \( \Thickness_{\Diffusivity{}} \) is the thickness of the
diffusion boundary layer.  The limiting deposition current \(
\Current_{\text{lim}} \) is determined by the complete depletion of \Cation{+} in
the electrolyte at the interface, such that
\begin{equation}
    \Current_{\text{lim}} 
    = \frac{
    	\Faraday\Diffusivity{\Cation{+}}
	\Concentration{\Cation{+}}^{\Thickness}}
    {\Thickness_{\Diffusivity{}}}.
    \label{eq:LimitingCurrent}
\end{equation}
The classical ``Butler-Volmer'' equation of electrochemistry is a special case of
Eq.~\eqref{eq:CurrentOverpotential} in which the effects of mass transfer are
neglected (\( \Current/\Current_{\text{lim}} \rightarrow 0 \)).

For small overpotentials, the linearized form of
Eq.~\eqref{eq:CurrentOverpotential} is
\begin{equation}
    \Overpotential \approx \Current\frac{\Gas\Temperature}{\Faraday}
    \left(
	\frac{1}{\Current_{0}} 
	- \frac{1}{\Current_{\text{lim}}}
    \right).
    \label{eq:CurrentOverpotential:Linear}
\end{equation}
We will use this relationship in Section~\ref{sec:Parameters:Interface} to relate
\( \Current_{0} \) to the parameters of our phase
field model.  When \( \abs{\Overpotential}\Faraday/\Gas\Temperature \gg
1 \) and \( \Current \ll \Current_{\text{lim}} \), Eq.~\eqref{eq:LimitingCurrent}
reduces to
\begin{subequations}
    \begin{align}
	\Current 
	&\approx \Current_{0} \exp
	\left[
	\frac{
	    \left(
		1-\Transfer
	    \right)
	    \Faraday\Overpotential}
	    {\Gas\Temperature}
	\right]
	\qquad\text{for \( \Overpotential > 0 \)}
	\label{eq:CurrentOverpotential:Tafel:Anodic} \\
\intertext{and}
	\Current 
	&\approx -\Current_{0} \exp
	\left[
	-\frac{\Transfer\Faraday\Overpotential}{\Gas\Temperature}
	\right]
	\qquad\text{for \( \Overpotential < 0 \).}
	\label{eq:CurrentOverpotential:Tafel:Cathodic}
    \end{align}%
    \label{eq:CurrentOverpotential:Tafel}%
\end{subequations}%
The quantities \( (1-\Transfer)\Faraday/\Gas\Temperature \)
and \( -\Transfer\Faraday/\Gas\Temperature \) are known as
the anodic and cathodic ``Tafel slopes'' from the slopes of the lines when \(
\ln\abs{\Current} \) is plotted against \Overpotential. These slopes can be 
used to deduce experimental values for \Transfer.

Eq.~\eqref{eq:CurrentOverpotential:General} was originally derived from reaction
rate theory to explain experimentally observed current-overpotential behavior.
More recently, atomistic and quantum mechanical treatments of electron and ion
transfer reactions have been performed to replace this chemical reaction rate
approach \cite{Schmickler:1996}.  These treatments have led to a better physical
understanding of the phenomenological constants \Transfer\ and \( \Current_{0}
\), but they do not fundamentally alter the form of
Eqs.~\eqref{eq:CurrentOverpotential:General} and \eqref{eq:CurrentOverpotential}.

\section{Model}
\label{sec:Model}

\subsection{General Kinetic Equations}
\label{sec:Model:Kinetics}

In Ref.~\cite{ElPhF:equilibrium}, we performed a variational analysis to
derive the governing equations for the equilibrium electrochemical interface. 
We also postulated the simplest time dependent forms of those governing
equations that guarantee a decrease in total free energy with time \Time.  We restate
those dynamic postulates here. The time variation of the phase field \Phase\ is 
given by
\begin{equation}
    \frac{\partial\Phase}{\partial\Time} 
    = -\Mobility{\Phase}\left[
	\frac{\partial\HelmholtzPerVol}{\partial\Phase}
	- \Gradient{\Phase}\nabla^2\Phase
	- \frac{\Dielectric'(\Phase)}{2} 
	    \left(\nabla\Potential\right)^{2}
    \right],
    \label{eq:Evolution:Phase}
\end{equation}
where \HelmholtzPerVol\ is the Helmholtz free energy density per unit volume,
\Gradient{\Phase} is the phase field gradient energy coefficient, \(
\Dielectric(\Phase) \) is the dielectric constant, \PRE{which we take to depend
explicitly on the phase; because all of the fields are coupled, it will also
depend implicitly on the electrolyte concentration}.  \Mobility{\Phase} is the
mobility of the phase field.  Under the assumption that all nonzero partial molar
volumes are identical, the flux \Flux{j} of each component \( j \) is
\begin{equation}
	\Flux{j} 
	= -\Mobility{j}\nabla\left[
	\Electrochemical{j} 
	    - \frac{\PartialMolarVolume{j}}
	    {\PartialMolarVolume{\Substitutional}}
	    \Electrochemical{\Solvent{}}
    \right],
    \qquad j = 1 \ldots \Components-1
    \label{eq:Flux}
\end{equation}
where \Electrochemical{j} is the electrochemical potential of species \( j \) and
\PartialMolarVolume{j} is the partial molar volume of species \( j \).  We divide
the components into electrons \Electron\ with \( j = 1 \), which have \(
\PartialMolarVolume{\Electron} = 0 \), and substitutional species with \( j > 1
\), which all have the same \( \PartialMolarVolume{j} =
\PartialMolarVolume{\Substitutional} = 0 \).  One consequence of this assumption
is that \( \sum_{j=2}^{\Components} \Concentration{j} =
\PartialMolarVolume{\Substitutional}^{-1} = \text{constant} \), where
\Concentration{j} is the concentration of species \( j \).  A specific choice is
made of a substitutional component \Solvent{} with nonzero partial molar volume
to be called the \PRE{reference species}.  The quantity \Mobility{j} is the mobility of component
\( j \).  Since conservation of species requires
\begin{equation}
    \frac{\partial\Concentration{j}}{\partial\Time}
    = - \nabla\cdot\Flux{j},
    \qquad j = 1 \ldots \Components-1
    \label{eq:Conservation:Concentration}
\end{equation}
one obtains
\begin{equation}
    \frac{\partial\Concentration{j}}{\partial\Time}
    = \nabla\cdot\left\{
	\Mobility{j}\nabla\left[
	    \Electrochemical{j} 
	    - \frac{\PartialMolarVolume{j}}
		{\PartialMolarVolume{\Substitutional}}
	    \Electrochemical{\Solvent{}}
	\right]
    \right\}.
    \qquad j = 1 \ldots \Components-1
    \label{eq:Evolution:Diffusion}
\end{equation}
Poisson's equation
\begin{equation}
    \nabla\cdot\left[\Dielectric(\Phase)\nabla\Potential\right] 
    + \ChargeDensity = 0
    \label{eq:Governing:Poisson}
\end{equation}
must also be satisfied everywhere, where the charge density is
\begin{equation}
    \ChargeDensity =
    \Faraday \sum_{j=1}^{\Components} \Valence{j} \Concentration{j}.
    \label{eq:ChargeDensity}
\end{equation}
The mobilities \Mobility{j} and \Mobility{\Phase} will be related to
the parameters of electrokinetics in
Sections~\ref{sec:Parameters:SinglePhase} and
\ref{sec:Parameters:Interface}.

\subsection{Form of the Dynamic Equations for Ideal Solution Behavior}

For simplicity, we assumed in Ref.~\cite{ElPhF:equilibrium} that the
chemical part of the Helmholtz free energy per unit volume is
described by an \PRE{interpolation between two} ideal solutions of the components,
\begin{multline}
    \HelmholtzPerVol\left(
    \Phase,\Concentration{j}
    \right)
    \\
    = \sum_{j=1}^{\Components} \Concentration{j} 
	\left\{
	\Chemical{j}{\Standard\Electrolyte}
	+ \Delta\Chemical{j}{\Standard}\Interpolate\left(\Phase\right)
	+ \Gas\Temperature\ln\Concentration{j}\MolarVolume
	+ \Barrier{j} \DoubleWell\left(\Phase\right)
	\right\},
	\label{eq:HelmholtzPerVolume}
\end{multline}
where the molar volume \( \MolarVolume = (\sum_{j=1}^{\Components}
\Concentration{j})^{-1} \).  We use an interpolating function \(
\Interpolate\left(\Phase\right) = \Phase^{3}\left(6\Phase^{2} - 15\Phase +
10\right) \) to bridge between the descriptions of the two bulk phases and a
double-well function \( \DoubleWell\left(\Phase\right) =
\Phase^{2}\left(1-\Phase\right)^{2} \) with a barrier height \Barrier{j} for each
component \( j \) to establish the metal/electrolyte interface \cite{Wang:1993}.
The polynomials are chosen to have the properties that \( \Interpolate(0) = 0 \),
\( \Interpolate(1) = 1 \), \( \Interpolate'(0) = \Interpolate'(1) = 0 \), and \(
\DoubleWell'(0) = \DoubleWell'(1) = 0 \).  The \PRE{classical} chemical potential
is given by \( \Chemical{j}{} =
\partial\HelmholtzPerVol/\partial\Concentration{j} \) and the corresponding
\PRE{classical} electrochemical potential is \( \Electrochemical{j} =
\Chemical{j}{} + \Valence{j}\Faraday\Potential \).

Substituting Eq.~\eqref{eq:HelmholtzPerVolume} into
Eqs.~\eqref{eq:Evolution:Phase} and \eqref{eq:Flux}, we obtain the governing
equation for evolution of the phase field under ideal solution thermodynamics
\begin{align}
    \frac{\partial\Phase}{\partial\Time} 
    &= -\Mobility{\Phase}\left[
	\Interpolate'\left(\Phase\right) 
	\sum_{j=1}^{\Components} \Concentration{j} \Delta\Chemical{j}{\Standard}
	+ \DoubleWell'\left(\Phase\right)
	\sum_{j=1}^{\Components} \Concentration{j} \Barrier{j}
    \right. \nonumber
    \\
    &\qquad\left.
	\vphantom{\sum_{j=1}^{\Components} \Concentration{j} \Delta\Chemical{j}{\Standard}}
	- \Gradient{\Phase}\nabla^2\Phase
	- \frac{\Dielectric'(\Phase)}{2} 
	    \left(\nabla\Potential\right)^{2}
    \right]
    \label{eq:Evolution:Ideal:Phase} 
\end{align}
and the flux in the diffusion equation Eq.~\eqref{eq:Conservation:Concentration}
is given by
\begin{subequations}
\begin{align}
	\Flux{j} 
    &= -\Mobility{j}\nabla\left[	
    	\left(\Delta\Chemical{j}{\Standard}-\Delta\Chemical{\Solvent{}}{\Standard}\right)
		\Interpolate\left(\Phase\right)
	+ \Gas\Temperature
	    \ln\frac{\Concentration{j}}{\Concentration{\Solvent{}}}
	\right. \nonumber
    \\
    &\qquad \qquad + \left. 
	\vphantom{\frac{\Concentration{j}}{\Concentration{\Solvent{}}}}
	\left(
	    \Valence{j} - \Valence{\Solvent{}}
	\right)\Faraday\Potential
	+ \left(
		\Barrier{j} - \Barrier{\Solvent{}}
	\right)\DoubleWell\left(\Phase\right)
    \right]
    \nonumber
    \\
    & \qquad \qquad \qquad j = 2 \ldots \Components-1
    \label{eq:Flux:Ideal:Substitutional}
    \\
	\Flux{\Electron} 
    &= -\Mobility{\Electron}\nabla\left[
	\Delta\Chemical{\Electron}{\Standard}\Interpolate\left(\Phase\right)
	+ \Gas\Temperature
	    \ln\frac{\PartialMolarVolume{\Substitutional}\Concentration{\Electron}}
		{1+\PartialMolarVolume{\Substitutional}\Concentration{\Electron}}
	\right. \nonumber
    \\
    &\qquad \qquad \left. 
	\vphantom{\frac{\PartialMolarVolume{\Substitutional}\Concentration{\Electron}}
		{1+\PartialMolarVolume{\Substitutional}\Concentration{\Electron}}}
	+ \Valence{\Electron}\Faraday\Potential
	+ \Barrier{\Electron} \DoubleWell\left(\Phase\right)
    \right].
    \label{eq:Flux:Ideal:Interstitial}    
\end{align}%
    \label{eq:Flux:Ideal}%
\end{subequations}%
The flux of substitutional species does not explicitly depend on the electron
concentration and the flux of electrons does not explicitly depend on the
concentration of substitional species; the flux of substitutional species is
affected by the displacement of other substitutional species, but electrons can
move without displacing other ions.  The fluxes of all species are coupled
indirectly through the total charge distribution and Eq.~\eqref{eq:Governing:Poisson}.

\section{Numerical Methods}
\label{sec:Numerics}

The 1-D form of the governing equations was transformed to a frame moving at a 
velocity \Velocity.
Simulations were performed in a domain of length \BoxSize\ with an initially
abrupt interface between the bulk electrode and electrolyte phases at \(
\Position = \BoxSize/2 \), such that \( \Phase|_{\Position<\BoxSize/2} =
\Phase^{\Electrode} = 1 \) and \( \Phase|_{\Position>\BoxSize/2} =
\Phase^{\Electrolyte} = 0 \).  After choosing an initial bulk value for \(
\Concentration{\Cation{+}}^{\Electrolyte} \); the remaining initial bulk \(
\Concentration{j}^{\Electrode} \) and \( \Concentration{j}^{\Electrolyte} \) were
the equilibrium values obtained by equating the bulk electrochemical potentials
\Electrochemical{j} \cite{ElPhF:equilibrium}.  The boundary condition on the
phase field is \( \vec{\Normal}\cdot\nabla\Phase = 0 \) at both ends of the
solution domain.  At the electrolyte end, we set \( \Potential = 0 \) and at the
electrode end we specify \( \Current \).  At the leading edge of the moving
frame, we model the stirred bulk electrolyte by applying a fixed concentration
boundary condition.  At the trailing edge of the frame, we discard the material
leaving the frame by setting the divergence of the species fluxes to zero.

Equations \eqref{eq:Conservation:Concentration}, \eqref{eq:Governing:Poisson},
\eqref{eq:Evolution:Ideal:Phase}, and \eqref{eq:Flux:Ideal} were solved with
explicit finite differences.  Spatial derivatives were taken to second order on a
uniform mesh.  Transient solutions were integrated numerically with an adaptive,
fifth-order Runge-Kutta time stepper (based on
\texttt{odeint} of Ref.~\cite{NumRec:C}) 
until a steady state was achieved (current became constant).  We have defined
steady state in our simulations as the point when each \( \Flux{j} - \Velocity
\Concentration{j} \) were uniform to within 0.1\%.  Because \Velocity\ is an
unknown result of the simulation, the frame velocity was adjusted at each
iteration to keep the interface stationary in the frame.

\section{Material Parameters}
\label{sec:Parameters}

\subsection{Equilibrium Material Parameters}
\label{sec:Parameters:Equilibrium}

We examine the dynamic behavior of a four component model under a different set
of thermodynamic parameters than described in Ref.~\cite{ElPhF:equilibrium}.  In
this paper, all components have valence \( \Valence{j} = \pm 1 \).  We are
primarily interested in the electrodeposition of the more noble cation
\Cation{+}, where the less noble cation \Otherion{+} and the anion \Anion{-} make
up the bulk of the supporting electrolyte.  This electrolyte containing only
charged species represents a molten salt system.  The presence of the second
cation \Otherion{+} introduces the possibility of alloy deposition.

We take the partial molar volume of the ``substitutional'' components
(\Cation{+}, \Otherion{+}, and \Anion{-}) as \(
\PartialMolarVolume{\Substitutional} =
\unit{\power{10}{-5}}{\meter\cubed\per\mole} \).  Equation~\eqref{eq:Equilibrium}
states that for any given \( \Fraction{j}^{\Electrode} \) and \(
\Fraction{j}^{\Electrolyte} \), there is some potential difference \Galvani{}
between that bulk phases that is related to the chemical potential difference of
the pure components \( \Delta\Chemical{j}{\Standard} \).  Conversely, we showed
in Ref.~\cite{ElPhF:equilibrium} that we can establish a value for \(
\Delta\Chemical{j}{\Standard} \) if we know \Galvani{} for some particular \(
\Fraction{j}^{\Electrode} \) and \( \Fraction{j}^{\Electrolyte} \), for instance
the standard state values
\begin{equation}
    \Delta\Chemical{j}{\Standard} 
    = \Gas\Temperature\ln
    \frac{\Fraction{j}^{\Electrolyte\Standard}}
	{\Fraction{j}^{\Electrode\Standard}}
    - \Valence{j}\Faraday\Galvani{\Standard}.
    \qquad j = 1\ldots\Components
    \label{eq:DeltaStandard:Standard}
\end{equation}
The voltage-independent
portion of \( \Delta\Chemical{j}{\Standard} \) is given in
Table~\ref{tab:Parameters:Thermodynamic:MoltenSalt}.
\begin{table}[tbp]
	\centering
	\caption{Numeric values of the potential-independent portion of the
	chemical potential differences \( \Delta\Chemical{j}{\Standard} \).}
    \begin{ruledtabular}
	\begin{tabular}{cd}
	    & \multicolumn{1}{c}
		{$\ln (\Fraction{j}^{\Electrolyte\Standard} /
		    \Fraction{j}^{\Electrode\Standard})$}
	\\

	\hline

	\Electron
	& -20.03
	\\
	
	\Cation{+}
	& -3.912
	\\
	
	\Otherion{+}
	& 20.01
	\\
	
	\Anion{-}
	& 20.03
	
    \end{tabular}
    \end{ruledtabular}    
	\protect\label{tab:Parameters:Thermodynamic:MoltenSalt}
\end{table}
In this paper, we take \Galvani{\Standard} to be zero.  Following
Ref.~\cite{ElPhF:equilibrium}, this implies that the equilibrium state for this
material system at the standard state concentration is near the point of zero
charge.  The mole fraction ratios in
Table~\ref{tab:Parameters:Thermodynamic:MoltenSalt} of the
normally-electroinactive species are chosen to give the corresponding small
standard state mole fractions as \( \Fraction{\Electron}^{\Electrolyte\Standard}
= \Fraction{\Otherion{+}}^{\Electrode\Standard} =
\Fraction{\Anion{-}}^{\Electrode\Standard} = \power{10}{-9} \).

\begin{figure}[tbp]
    \centering
    \includegraphics[width=\figurewidth]{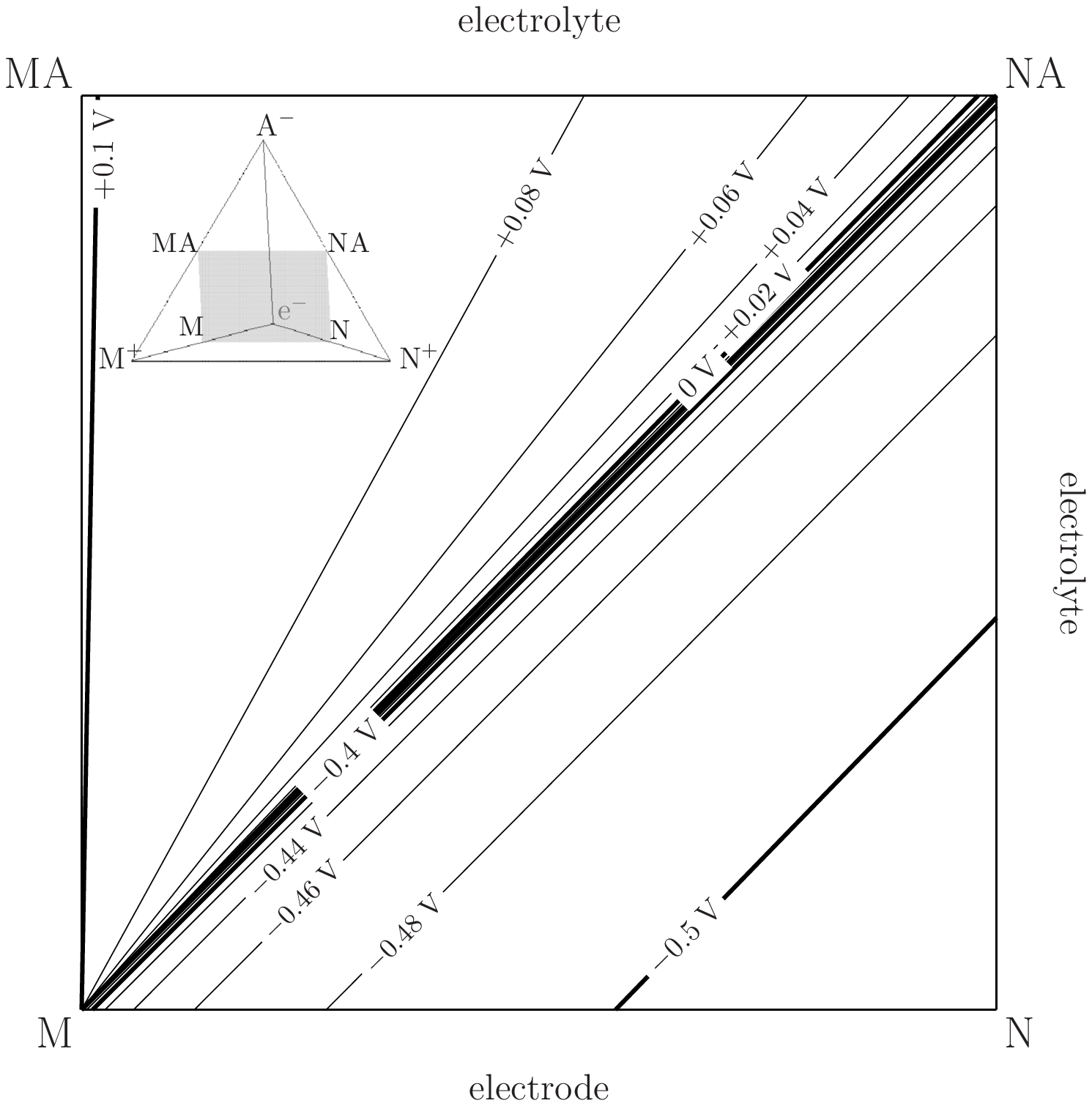}
    \caption{\PRE{Potential-composition} phase diagram for the parameters in
    Table~\ref{tab:Parameters:Thermodynamic:MoltenSalt}, \PRE{illustrating the
    bulk equilibrium between a \Cation{} electrode and an electrolyte containing
    \Cation{}\Anion{} salt dissolved in \Otherion{}\Anion{}}.  Tie-lines denote
    different values of the quantity \( (\Galvani{} - \Galvani{\Standard}) \).
    The inset shows the position of this charge neutral phase diagram within the
    quaternary domain of the charged species.}
    \protect\label{fig:Nernst}
\end{figure}
To permit a convenient graphical display of bulk equilibrium, we invoke charge
neutrality to transform the four components \{\Electron, \Cation{+},
\Otherion{+}, \Anion{-}\} into an alternate set of four components that are
charge-neutral, \{\Cation{}, \Otherion{}, \Cation{}\Anion{},
\Otherion{}\Anion{}\}.  We plot the equilibrium phase diagram in terms of these
transformed components in Figure~\ref{fig:Nernst}.  Equilibrium states exist only
between \( \unit{-0.5138}{\volt} < \Galvani{} < \unit{+0.1005}{\volt} \).  It can
be seen that over the majority of the potential range, from \( \unit{-0.4}{\volt}
\lesssim \Galvani{} \lesssim \unit{0}{\volt} \), that the equilibrium is between
an electrode of essentially pure \Cation{} and a \Otherion{}\Anion{} electrolyte
containing a dilute concentration of \Cation{}\Anion{}.  At the positive
\Galvani{} extreme, the equilibrium is between \Cation{} and \Cation{}\Anion{}
and at the negative \Galvani{} extreme, the equilibrium is between a \PRE{phase} of
\Cation{} and \Otherion{} and a \PRE{solution of} \Otherion{} and \Otherion{}\Anion{} for
this choice of \( \Fraction{j}^{\Standard} \).

Table~\ref{tab:Parameters:Interface} lists our choice of the parameters that
characterize the thickness and energy of the electrode-electrolyte interface.
\begin{table}[tbp]
    \caption{Parameters characterizing the equilibrium interface. \PermittivityVacuum\ 
    is the permittivity of free space.}
    \begin{ruledtabular}    
    \begin{tabular}{cc}
	parameter & value \\
	\hline
	\Gradient{\Phase} 
	& \unit{\Scientific{7.2}{-11}}{\joule\per\meter}  \\
	\Barrier{j\in 2\ldots\Components} 
	& \unit{\Scientific{3.6}{5}}{\joulepermole}  \\
	\Barrier{\Electron{}} 
	& \unit{0}{\joulepermole}  \\
	\Dielectric
	& 8\PermittivityVacuum
    \end{tabular}
    \end{ruledtabular}    
    \protect\label{tab:Parameters:Interface}
\end{table}
Our assumption that the barrier heights \Barrier{j} are equal for the
substitutional species and zero for electrons is discussed in
Ref.~\cite{ElPhF:equilibrium}.

\subsection{Single Phase Transport Properties (values for \Mobility{j})} 
\label{sec:Parameters:SinglePhase}

To identify the mobilities \Mobility{j}, we examine single-phase
systems.  In a single-phase electrode, \( \Phase =
\Interpolate\left(\Phase\right) = 1 \).  In a single-phase electrolyte,
\( \Phase = \Interpolate\left(\Phase\right) = 0 \).  In either phase
\( \DoubleWell\left(\Phase\right) =
\nabla\Interpolate\left(\Phase\right) =
\nabla\DoubleWell\left(\Phase\right) = 0 \).  We thus can write the
fluxes in Eq.~\eqref{eq:Flux:Ideal} as
\begin{subequations}
\begin{align}
    \Flux{j}^\text{bulk}
    &= -\Mobility{j}\left[	
	\frac{\Gas\Temperature\left(
	    \Concentration{\Solvent{}}+\Concentration{j}
	\right)}{\Concentration{\Solvent{}}\Concentration{j}}
	\nabla\Concentration{j}
	+ \frac{\Gas\Temperature}{\Concentration{\Solvent{}}}
	\sum_{\substack{i=2\\ i \neq j}}^{\Components-1} \nabla\Concentration{i}
	\right. \nonumber 
	\\
	&\qquad \left.
	\vphantom{\sum_{\substack{i=2\\ i \neq j}}^{\Components-1} \nabla\Concentration{i}}
	+ \left(
	    \Valence{j} - \Valence{\Solvent{}}
	\right)\Faraday\nabla\Potential
    \right], \qquad j = 2 \ldots \Components-1
    \label{eq:Flux:Bulk:Substitutional}
    \\
    \Flux{\Electron}^\text{bulk} 
    &= -\Mobility{\Electron}\left[
	    \frac{\Gas\Temperature\nabla\Concentration{\Electron}}{
	    \left(
		1+\PartialMolarVolume{\Substitutional}\Concentration{\Electron}
	    \right)\Concentration{\Electron}}
	+ \Valence{\Electron}\Faraday\nabla\Potential
    \right].
    \label{eq:Flux:Bulk:Interstitial}    
\end{align}
    \label{eq:Flux:Bulk}    
\end{subequations}
The total current is given by the relationship
\begin{equation}
    \vec{\Current} 
    = \Faraday \sum_{j=1}^{\Components} \Valence{j}\Flux{j}.
    \label{eq:Current}
\end{equation}
The flux of component \Components\ balances the other fluxes such that
\begin{equation}
    \sum_{j=2}^{\Components} \Flux{j} = 0.
    \label{eq:Flux:Substitutional:Constraint}
\end{equation}

We first consider an electrolyte with \( \nabla\Potential = 0 \).  If we
compare the resulting form of Eq.~\eqref{eq:Flux:Bulk:Substitutional}
with the classical diffusive flux equation with diffusivities 
\Diffusivity{ij},
\begin{equation}
    \Flux{j}
	= -\sum_{i=2}^{\Components-1}
    \Diffusivity{ij} \nabla \Concentration{i},
    \qquad j = 2 \ldots \Components-1
    \label{eq:Diffusion:Classical}
\end{equation}
the mobilities can be expressed in terms of the diagonal elements of 
\Diffusivity{ij} as
\begin{equation}
    \Mobility{j} 
    = \frac{\Diffusivity{jj}\Concentration{\Solvent{}}\Concentration{j}}
		{\Gas\Temperature
	\left(\Concentration{\Solvent{}}+\Concentration{j}\right)}.
	\qquad j = 2 \ldots \Components-1
	\label{eq:Mobility:Subsitutional}
\end{equation}
For simplicity, we assume the diagonal elements of \Diffusivity{ij} are
constants, thus inducing a concentration dependence in the mobilities as defined
by Eq.~\eqref{eq:Mobility:Subsitutional} and in the off-diagonal 
\Diffusivity{ij}'s.

We next consider an electrode with all \( \nabla\Concentration{j} = 0
\), where the current is entirely carried by the electromigration of
electrons.  The resulting form of
Eq.~\eqref{eq:Flux:Bulk:Interstitial}
\begin{equation}
    \Flux{\Electron} 
    = -\Mobility{\Electron}\Valence{\Electron}\Faraday\nabla\Potential
    \label{eq:Electromigration:Electron}
\end{equation}
can be substituted into Eq.~\eqref{eq:Current} to give
\begin{equation}
	\vec{\Current} 
    \approx -\Valence{\Electron}^{2} \Faraday^{2} \Mobility{\Electron}
	\nabla\Potential
    \label{eq:Current:Approximate}
\end{equation}
By comparison with Ohm's law, \( \vec{\Current} =
-\Conductivity\nabla\Potential \), we readily see that the
electron mobility
\begin{equation}
    \Mobility{\Electron}
    = \frac{\Conductivity}{\Valence{\Electron}^{2} \Faraday^{2}}
    = \frac{\Conductivity}{\Faraday^{2}}.
    \label{eq:Conductivity}
\end{equation} 
Thus Eqs.~\eqref{eq:Mobility:Subsitutional} and
\eqref{eq:Conductivity} relate the \Mobility{j}'s to the electronic
conductivity and ionic diffusivities.

On substitution of Eq.~\eqref{eq:Mobility:Subsitutional} into
Eq.~\eqref{eq:Flux:Bulk:Substitutional}, we see that the electromigration flux
(due to gradients in \Potential) within the electrolyte is
\begin{align}
    \Flux{j}^\Potential
    &= -\Mobility{j}\left(\Valence{j}-\Valence{\Solvent{}}\right)
    \Faraday \nabla \Potential
    \nonumber \\
    &= -\frac{\Diffusivity{jj}\left(\Valence{j}-\Valence{\Solvent{}}\right)
		\Concentration{\Solvent{}}\Concentration{j}\Faraday}
	{\Gas\Temperature\left(
	    \Concentration{\Solvent{}}+\Concentration{j}
	\right)}\nabla\Potential
    \nonumber \\
    &\approx 
-\frac{\Diffusivity{jj}\left(\Valence{j}-\Valence{\Solvent{}}\right)
		\Concentration{j}\Faraday}
	{\Gas\Temperature} \nabla \Potential.
	\qquad j = 2 \ldots \Components-1
    \label{eq:electromigration}
\end{align}
This is just as expected from traditional electrochemical theory, in the dilute
limit where \( \Concentration{\Solvent{}}/(\Concentration{\Solvent{}} +
\Concentration{j}) \approx 1 \).  We will find in
Section~\ref{sec:Solutions:SinglePhase} that, for our supported ionic
electrolyte and our electronic conducting electrode, we are justified
in neglecting the contributions of the electromigration current in the
bulk electrolyte and of the diffusion current in the bulk electrode.

It is interesting to note that the conductivity predicted by 
Eq.~\eqref{eq:Conductivity}
is completely analogous to that predicted by the Drude model (and by
the Fermi-Dirac model, for that matter) \cite{AshcroftMermin}
\begin{equation}
    \Conductivity =
    \frac{\Valence{\Electron}^{2} \Faraday^{2} \tau
    \Concentration{\Electron}}{\Mass_{\Electron}},
    \label{eq:Conductivity:Drude}
\end{equation}
where \( \Mass_{\Electron} \) is the mass of the electron.  The relaxation time
\( \tau \) can only be determined by quantum mechanical means and is simply an
unknown constant in classical models of electron transport.  Following an
analysis for the electrons similar to that which gave us
Eq.~\eqref{eq:Mobility:Subsitutional}, we find that we can describe the mobility
of electrons \Mobility{\Electron} in terms of a constant diffusivity of
electrons \Diffusivity{\Electron}
\begin{equation}
	\Mobility{\Electron} 
    = \frac{\Diffusivity{\Electron}\left(
		1+\PartialMolarVolume{\Substitutional}\Concentration{\Electron}
	    \right)\Concentration{\Electron}}
		{\Gas\Temperature}.
	\label{eq:Mobility:Interstitial}
\end{equation}
In a single-phase conductor with uniform concentrations, \( (1 +
\PartialMolarVolume{\Substitutional}\Concentration{\Electron}) \) is a
dimensionless constant of order 1.  We can see that \( \Diffusivity{\Electron}
/ (\Gas\Temperature) \) is dimensionally equivalent to \( \tau /
\Mass_{\Electron} \) and all other terms in Eqs.~\eqref{eq:Conductivity} and
\eqref{eq:Conductivity:Drude} are identical.  The room temperature conductivity
of silver of approximately
\unit{\Scientific{6}{7}}{\reciprocal\ohm\reciprocal\meter} results in \(
\Diffusivity{\Electron} \approx \unit{\Scientific{8}{-5}}{\squaremetrepersecond}
\) and \( \Mobility{\Electron} \approx \unit{\Scientific{6}{-3}}
{\mole\squared\per(\joule.\second.\meter)} \).  We observe that one
of the weaknesses of the Drude model is that it fails to predict the \(
\Conductivity\sim\Temperature^{-1} \) dependence found in experiments without
making some unsatisfactory ad hoc assumptions; this dependence arises naturally
in our fundamentally thermodynamic formulation.

\subsection{Interfacial Kinetics (value for \Mobility{\Phase})}
\label{sec:Parameters:Interface}

\begin{figure*}[tbp]
    \centering
    \subfigure{\label{fig:LinearKinetics:Dimensional}%
	\includegraphics[width=\subfigurewidth]{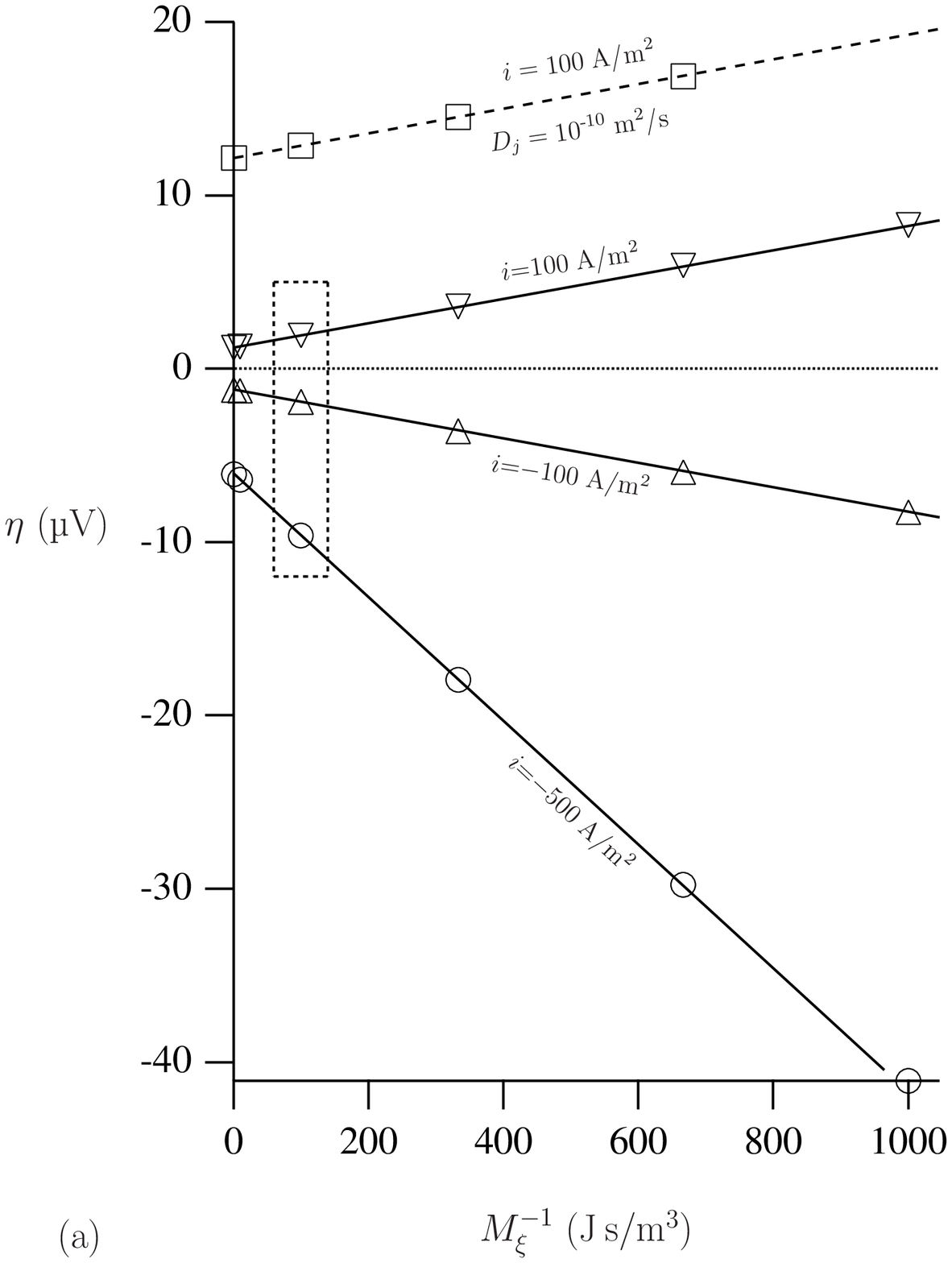}%
	}%
    \subfigure{\label{fig:LinearKinetics:Dimensionless}%
	\includegraphics[width=\subfigurewidth]{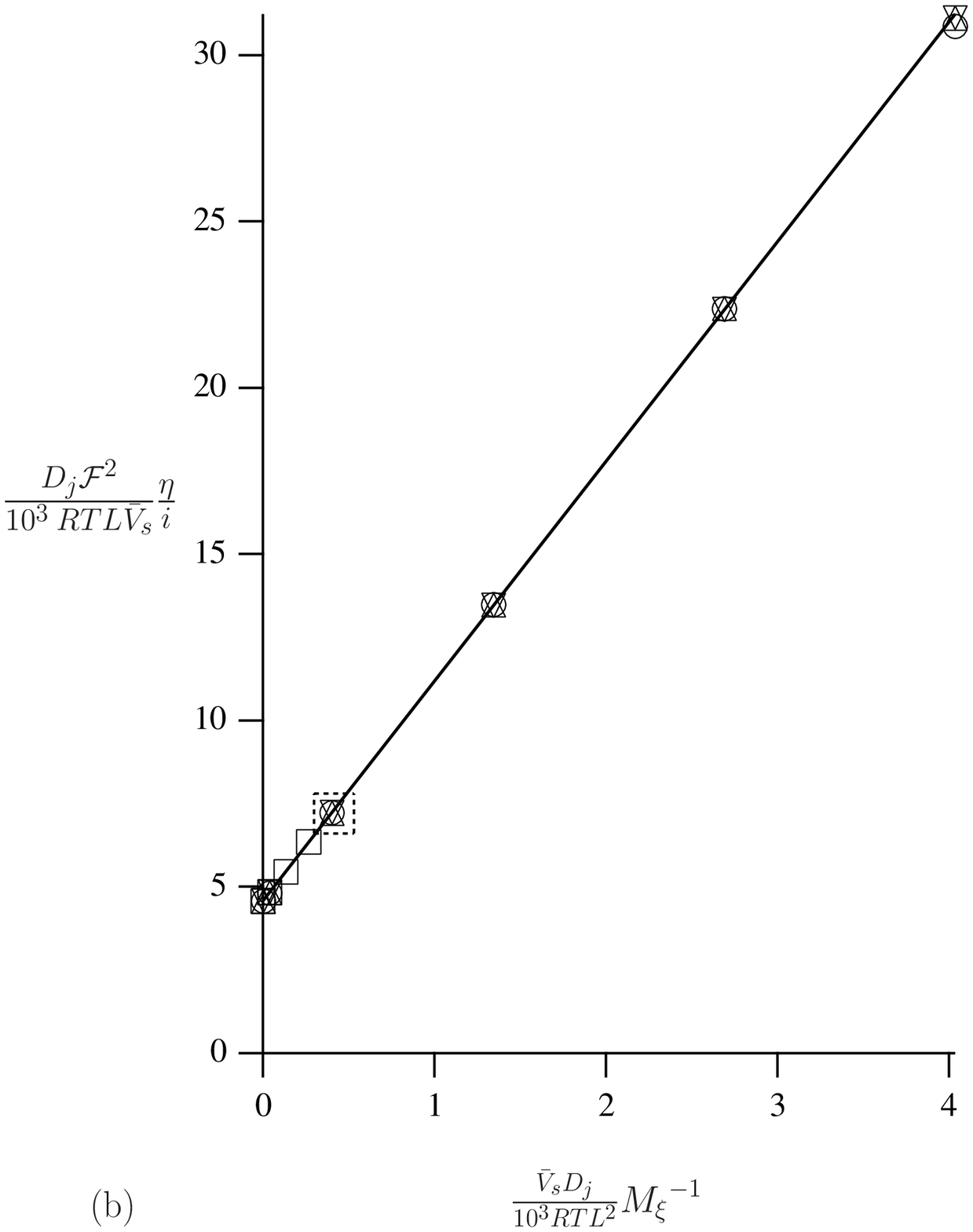}%
	}%
    \caption{\PRE{(a) Dimensional and (b) dimensionless} relationship between
    overpotential \Overpotential\ and current \Current\ as a function of the
    inverse phase field mobility \( {\Mobility{\Phase}}^{-1} \) when \(
    \Concentration{\Cation{+}}^{\Electrolyte} = \unit{10}{\mole\per\meter\cubed}
    \) and \( \Diffusivity{jj} = \unit{\power{10}{-9}}{\squaremetrepersecond} \).
    Points are plotted for each permutation of \( \Current = \unit{(-500, -100,
    100)}{\amperepersquaremetre} \) and \( \Mobility{\Phase} =
    \unit{(\power{10}{-3}, \Scientific{1.5}{-3}, \Scientific{3}{-3},
    \power{10}{-2}, \power{10}{-1}, 1)}{\meter\cubed\per(\joule.\second)} \).
    Points are also plotted for \( \Current = \unit{100}{\amperepersquaremetre}
    \) and \( \Mobility{\Phase} = \unit{(\Scientific{1.5}{-3},
    \Scientific{3}{-3}, \power{10}{-2}, 1)}{\meter\cubed\per(\joule.\second)} \)
    with \( \Diffusivity{jj} = \unit{\power{10}{-10}}{\squaremetrepersecond} \).
    The line in Figure~(b) is a fit to \( \Overpotential/\Current =
    a{\Mobility{\Phase}}^{-1}+b \) with \( a = (6.610 \pm 0.006)
    (\PartialMolarVolume{\Substitutional}^{2}/\BoxSize\Faraday^{2}) \) and \( b =
    (4564 \pm 1) (\Gas\Temperature\PartialMolarVolume{\Substitutional}\BoxSize
    /\Diffusivity{jj}\Faraday^{2}) \).  The points in the dotted box contribute
    to Figure~\ref{fig:CurrentOverpotential}.}
    \protect\label{fig:LinearKinetics}
\end{figure*}

Along with the transfer coefficient \Transfer, the exchange current \(
\Current_{0} \) characterizes the kinetics of the interface and we hypothesize
that it has an intimate relationship to the phase field mobility
\Mobility{\Phase}.  To test this hypothesis for our model, we 
examine Eq.~\eqref{eq:CurrentOverpotential:Linear} and plot
\Overpotential\ obtained from steady-state calculations against \(
{\Mobility{\Phase}}^{-1} \) for various \Diffusivity{jj} and small values of
\Current\ in Figure~\ref{fig:LinearKinetics}.
If we scale length by \BoxSize\ (the length of the solution domain), time by \(
\LengthScale^{2}/\Diffusivity{jj} \), energy density by \(
\Gas\Temperature/\PartialMolarVolume{\Substitutional} \), and potential by \(
\Gas\Temperature/\Faraday \), we find that all of the points satisfy the linear
relationship
\begin{equation}
    \Overpotential / \Current
    = (6.610 \pm 0.006){\Mobility{\Phase}}^{-1} 
    \frac{\PartialMolarVolume{\Substitutional}^{2}}{\BoxSize\Faraday^{2}}
    + (4563 \pm 1)
    \frac{\Gas\Temperature\PartialMolarVolume{\Substitutional}\BoxSize}
	{\Diffusivity{jj}\Faraday^{2}},
    \label{eq:CurrentOverpotential:Linear:Fit}
\end{equation}
Comparison with Eq.~\eqref{eq:CurrentOverpotential:Linear} reveals
that
\begin{align}
    \Current_{0} 
    &= (0.1513 \pm 0.0001) \Mobility{\Phase} \frac{\Gas\Temperature\Faraday\BoxSize}
    {\PartialMolarVolume{\Substitutional}^{2}}
    \nonumber \\
    &\approx \unit{\Scientific{3.62}{6}}{\amperepersquaremetre}
    \qquad\text{when \( \Mobility{\Phase} =
    \unit{\power{10}{-2}}{\meter\cubed\per(\joule.\second)} \)}
    \label{eq:ExchangeCurrent:Fit}
\end{align}
and 
\begin{align}
    \Current_{\text{lim}}
    &= \Scientific{(2.191 \pm 0.0004)}{-4}
    \frac{\Diffusivity{jj}\Faraday}{\BoxSize\PartialMolarVolume{\Substitutional}}.
    \nonumber \\
    &\approx \unit{\Scientific{2.11}{6}}{\amperepersquaremetre}
    \qquad\text{when \( \Diffusivity{jj} =
    \unit{\power{10}{-9}}{\squaremetrepersecond} \)}
    \label{eq:LimitingCurrent:Fit}
\end{align}
Eq.~\eqref{eq:ExchangeCurrent:Fit} confirms our hypothesis that \( \Current_{0}
\) is directly related to \Mobility{\Phase}.  Comparing
Eq.~\eqref{eq:LimitingCurrent:Fit} to Eq.~\eqref{eq:LimitingCurrent}, and taking
\( \Concentration{\Cation{+}}^{\Thickness} = \unit{10}{\mole\per\meter\cubed} \),
we see that this implies that the diffusion boundary layer thickness is \(
\Thickness_{\Diffusivity{}} = (0.4564 \pm 0.0001) \BoxSize \).  This is very close
to the thickness of the electrolyte, which validates that we are computing the
diffusion field correctly (because we are modeling a diffuse interface, the
electrolyte thickness is somewhat less than \( 0.5 \BoxSize \)).  The thinness of
the diffusion boundary layer in our calculations gives rise to a limiting current
that is much larger than encountered in physical systems, but the mechanism is
the same.

\begin{table*}[tbp]
	\centering
	\caption{Correspondence between kinetic parameters used in this phase
	field model and those measured in experiments or typical of
	sharp-interface models.  Typical physical values \cite{Bard:2nd} are
	compared with the values used in our numerical calculations.  The
	diffusivities are given for the electrolyte phase; diffusivities in the
	solid electrode are expected to be many orders of magnitude smaller.  For
	the calculations in this paper, we treat the diagonal diffusivities as
	constant and uniform.  To simplify the notation, we take \(
	\Diffusivity{j} \equiv \Diffusivity{jj} \).  No \Diffusivity{\Anion{-}}
	is necessary because \Anion{-} is the \PRE{reference} species in our
	calculations.}
    \begin{ruledtabular}
	\begin{tabular}{@{}===@{}}
	    \multicolumn{1}{c}{phase field}
	    & \multicolumn{1}{c}{``physical''}
	    & \multicolumn{1}{c}{``numeric''}
	\\ 
	
	\hline
		
	\Diffusivity{\Electron}
	    = \unit{\power{10}{-9}}{\squaremetrepersecond}
	& \Conductivity
		= \unit{\Scientific{6}{7}}
	    {\reciprocal\ohm\reciprocal\meter}
	& \Conductivity
		= \unit{750}
	    {\reciprocal\ohm\reciprocal\meter}
	\\
		
	\Diffusivity{\Cation{+}}
	= \unit{\power{10}{-9}}{\squaremetrepersecond}
	& \Diffusivity{\Cation{+}}
		= \unit{\power{10}{-9}}{\squaremetrepersecond}
	& \Diffusivity{\Cation{+}}
		= \unit{\power{10}{-9}}{\squaremetrepersecond}
	\\

	\Diffusivity{\Otherion{+}}
	= \unit{\power{10}{-9}}{\squaremetrepersecond}
	& \Diffusivity{\Otherion{+}}
		= \unit{\power{10}{-9}}{\squaremetrepersecond}
	& \Diffusivity{\Otherion{+}}
		= \unit{\power{10}{-9}}{\squaremetrepersecond}

	\\

	\Mobility{\Phase}
	= \unit{\power{10}{-2}}{\meter\cubed\per(\joule.\second)}
	& 
	\Current_{0}
	= \unit{(\text{\power{10}{-16} to \power{10}{-2}})}{\amperepersquaremetre}
	& \Current_{0}
	= \unit{\Scientific{3.7}{6}}{\amperepersquaremetre}
    \end{tabular}
    \end{ruledtabular}    
	\protect\label{tab:Parameters:Kinetic}
\end{table*}
Table~\ref{tab:Parameters:Kinetic} displays the kinetic parameters of the phase
field model and typical values of the corresponding physical quantities.  If
physical values are used for some kinetic parameters, then the computation time is
too long, so the values used for our numeric simulations are also listed. 

\section{Numerical Results and Discussion}
\label{sec:Solutions}

Our purpose is to show consistency of behavior with sharp interface models of
electrochemical systems so that future 2-D and 3-D computations treating more
complex phenomena can be performed with confidence.  In this section, we examine
the behavior of our model in the bulk phases, explore the current-overpotential
behavior, and demonstrate the electrodeposition of alloys at high applied
currents.

\begin{figure}[tbp]
    \centering
    \includegraphics[width=\figurewidth]{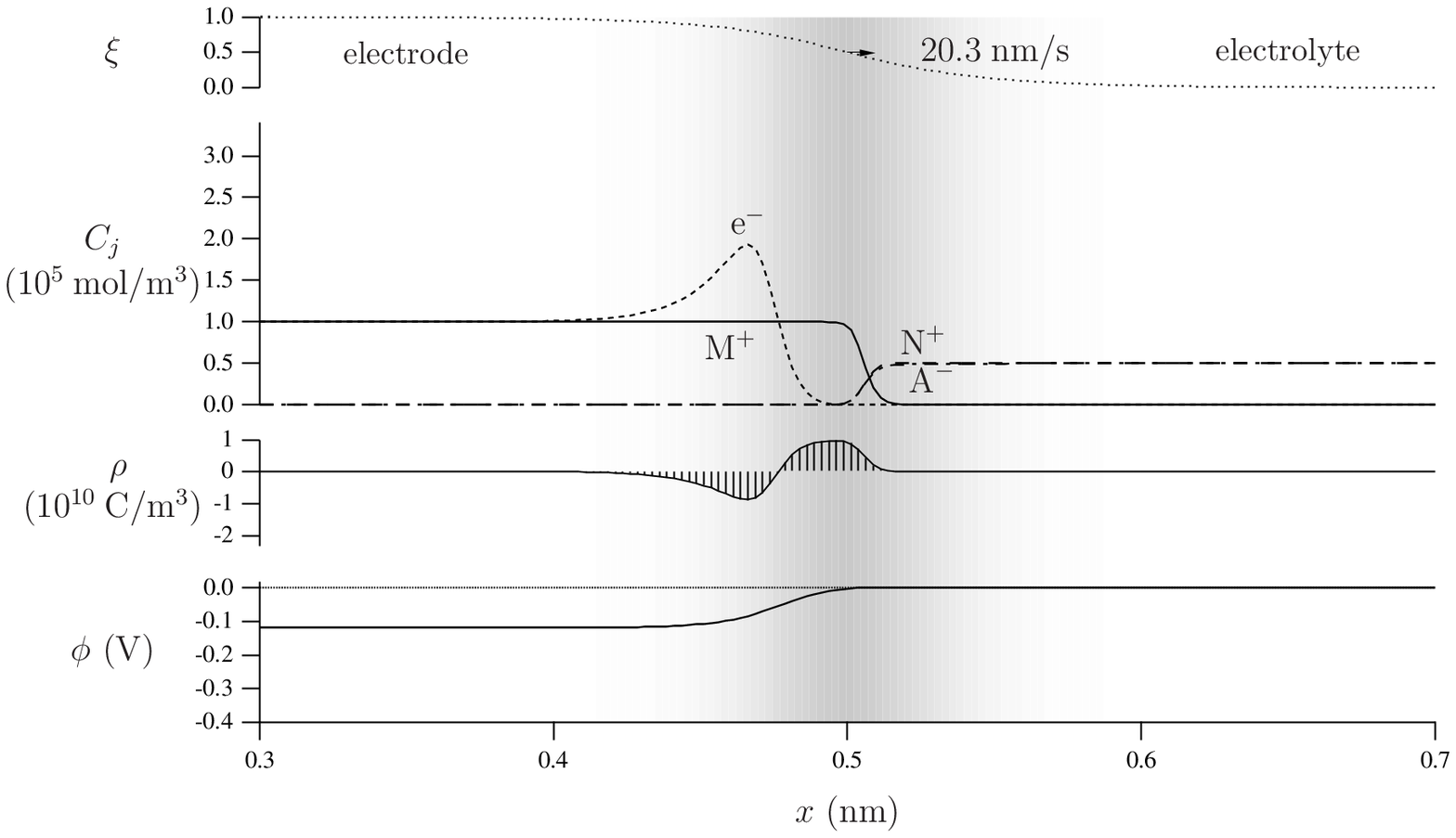}
    \caption{Interface profiles for steady state electrodeposition with \(
    \Current = \unit{\power{-10}{2}}{\amperepersquaremetre} \).  The
    concentration profiles for \Otherion{+} and \Anion{-} are almost coincident
    on this scale.  \PRE{\( \DoubleWell(\Phase) \) is mapped onto the background in
    gray to indicate the location of the phase field interface.}}
    \protect\label{fig:SteadyState}
\end{figure}
The interfacial region of a representative steady-state solution, with \(
\Current = \unit{-100}{\amperepersquaremetre} \), is displayed in
Figure~\ref{fig:SteadyState}.  The phase field \Phase, concentrations
\Concentration{j}, charge density \ChargeDensity, and electrostatic potential
\Potential\ are plotted against the same \Position-axis.  The velocity of the
moving frame is indicated with a marker on the \Phase\ curve at \( \Phase = 0.5
\).  To highlight the location of the interface, \( \DoubleWell(\Phase) \) is
mapped onto the background in gray.  We can see that the concentrations deviate
from their bulk values in a region of approximately the same thickness as the
phase field transition.  As a result, the charged ``double layer'' is confined to
this same region.  The surface of the electrode has excess \Electron, whereas the
surface of the electrolyte is an essentially charge-neutral \Otherion{}\Anion{}
salt with a dilute concentration of \Cation{}\Anion{}.  All of the species except
\Cation{+} are excluded from the region of intermediate \Phase, giving rise to a
layer of \Cation{+} that has neither \Electron\ nor \Anion{-} to balance the
charge.  This charge distribution gives rise to the potential step of
approximately \unit{0.12}{\volt} between the two phases, which is the expected
Nernst potential of an electrolyte with \(
\Concentration{\Cation{+}}^{\Electrolyte} = \unit{10}{\mole\per\meter\cubed} \).

\subsection{Fluxes}
\label{sec:Solutions:SinglePhase}

\begin{table*}[tbp]
    \caption{Partial fluxes in the bulk electrolyte for \( 
    \Current = \unit{-100}{\amperepersquaremetre} \) (electrodeposition). \( \nabla\Potential = 
    \unit{\Scientific{6.87}{-4}}{\voltpermetre} \).}
    \protect\label{tab:Bulk:Electrolyte}
    \begin{ruledtabular}
    \begin{tabular}{cxxxxx}
	
	\( j \) 
	& \multicolumn{1}{c}{\Concentration{j}\per(\mole\per\meter\cubed)}
	& \multicolumn{1}{c}{\( \nabla\Concentration{j}\per(\mole\per\power{\meter}{4}) \)}
	& \multicolumn{1}{c}{\( \Flux{j}^{\Diffusivity{}}\per(\mole\usk\meter\rpsquared\usk\reciprocal\second) \)}
	& \multicolumn{1}{c}{\( \Flux{j}^{\Potential}\per(\mole\usk\meter\rpsquared\usk\reciprocal\second) \)}
	& \multicolumn{1}{c}{\( \Flux{j}^\text{total}\per(\mole\usk\meter\rpsquared\usk\reciprocal\second) \)}
	\\
	
	\hline
	
	\Electron &  \SciTab{1.00}{-2} & \SciTab{-1.65}{3} 
	& \SciTab{1.65}{-6} & 0 & \SciTab{1.65}{-6}  \\
	
	\Cation{+} & \SciTab{1.00}{1} & \SciTab{1.03}{6} 
	& \SciTab{-1.03}{-3} & 0 & \SciTab{-1.03}{-3}  \\
	
	\Otherion{+} & \SciTab{5.00}{4} & \SciTab{-1.03}{6} 
	& \SciTab{5.18}{-4} & \SciTab{-1.34}{-6} & \SciTab{5.16}{-4}  \\

	\Anion{-} & \SciTab{5.00}{4} & \SciTab{8.28}{2} 
	& \multicolumn{1}{c}{---} & \multicolumn{1}{c}{---} & \SciTab{5.17}{-4} 
	
    \end{tabular}
    \end{ruledtabular}
\end{table*}
\begin{table*}[tbp]
    \caption{Partial fluxes in the bulk electrode for \( 
    \Current = \unit{-100}{\amperepersquaremetre} \) (electrodeposition). \( \nabla\Potential = 
    \unit{0.133}{\voltpermetre} \).}
    \protect\label{tab:Bulk:Electrode}
    \begin{ruledtabular}
    \begin{tabular}{cxxxxx}
	
	\( j \) 
	& \multicolumn{1}{c}{\Concentration{j}\per(\mole\per\meter\cubed)}
	& \multicolumn{1}{c}{\( \nabla\Concentration{j}\per(\mole\per\power{\meter}{4}) \)}
	& \multicolumn{1}{c}{\( \Flux{j}^{\Diffusivity{}}\per(\mole\usk\meter\rpsquared\usk\reciprocal\second) \)}
	& \multicolumn{1}{c}{\( \Flux{j}^{\Potential}\per(\mole\usk\meter\rpsquared\usk\reciprocal\second) \)}
	& \multicolumn{1}{c}{\( \Flux{j}^\text{total}\per(\mole\usk\meter\rpsquared\usk\reciprocal\second) \)}
	\\
	
	\hline
	
	\Electron & \SciTab{1.00}{5} & \SciTab{4.97}{2} 
	& \SciTab{-4.97}{-7} & \SciTab{1.03}{-3} & \SciTab{1.03}{-3}  \\
	
	\Cation{+} & \SciTab{1.00}{5} & \SciTab{-5.93}{5} 
	& \SciTab{1.85}{-7} & 0 &  \SciTab{1.85}{-7}  \\
	
	\Otherion{+} & \SciTab{2.04}{-2} & \SciTab{5.93}{5} 
	& \SciTab{1.24}{-7} & 0 & \SciTab{1.24}{-7} \\

	\Anion{-} & \SciTab{2.10}{-6} & \SciTab{1.88}{2} 
	& \multicolumn{1}{c}{---} & \multicolumn{1}{c}{---} & \SciTab{-3.10}{-7}
	
    \end{tabular}
    \end{ruledtabular}
\end{table*}
The relative contributions of the flux due to diffusion \(
\Flux{j}^{\Diffusivity{}} \) (dependent on all the \( \nabla\Concentration{i} \))
and the flux due to electromigration \( \Flux{j}^{\Potential} \) (proportional to
\( \nabla\Potential \)) can be distinguished using Eq.~\eqref{eq:Flux:Bulk}.  For
\( \Current = \unit{-100}{\amperepersquaremetre} \), the partial fluxes in the
bulk electrolyte are listed in Table~\ref{tab:Bulk:Electrolyte} and those in the
bulk electrode are listed in Table~\ref{tab:Bulk:Electrode}.  As the designated
\PRE{reference species}, the flux of \Anion{-} always adjusts such that the sum
of the fluxes of the substitutional species is zero.  In both phases, the
concentration gradients of \Cation{+} and \Otherion{+} are approximately equal
and opposite in sign to maintain charge neutrality (the concentration gradients
of \Anion{-} and \Electron\ are small).  The diffusive fluxes of \Cation{+} and
\Otherion{+} are \emph{not} equal and opposite in sign.  The ``off-diagonal''
term for the \Otherion{+} flux in the electrode and for both the \Cation{+} and
\Otherion{+} fluxes in the electrolyte contribute significantly.

Because we consider a supported electrolyte (the total ion density is high), \(
\nabla\Potential \) is small and electromigration does not contribute
significantly to the current in the electrolyte.  Both the magnitude and gradient
of \Concentration{\Electron} are small in the electrolyte, such that \Electron\
do not carry any significant current in the electrolyte.  The current due to the
\Otherion{+} flux is cancelled by that due to the \Anion{-} flux, such that
essentially all of the current in the electrolyte is carried by the diffusion of
\Cation{+}.  In the electrode, the partial fluxes of the substitutional
components are numerically zero.  The concentration gradient of \Electron\ is
small in the electrode, giving a small diffusive flux.  The bulk of the current
in the electrode is carried by electromigration of \Electron, consistent with
Ohm's law.  These observations that the current in the electrolyte is carried by
diffusion of \Cation{+} and the current in the electrode is carried by the
electromigration of \Electron\ are consistent with the approximations we made for
the bulk phases in Section~\ref{sec:Parameters:SinglePhase}; \emph{i.e.}, bulk
behavior is obtained at a distance of \unit{0.5}{\nano\meter} from the interface.

\subsection{Diffusion Layer}

\begin{figure*}[tbp]
    \centering
    \subfigure{\label{fig:concOverpotential:Plating}%
	\includegraphics[width=\subfigurewidth]{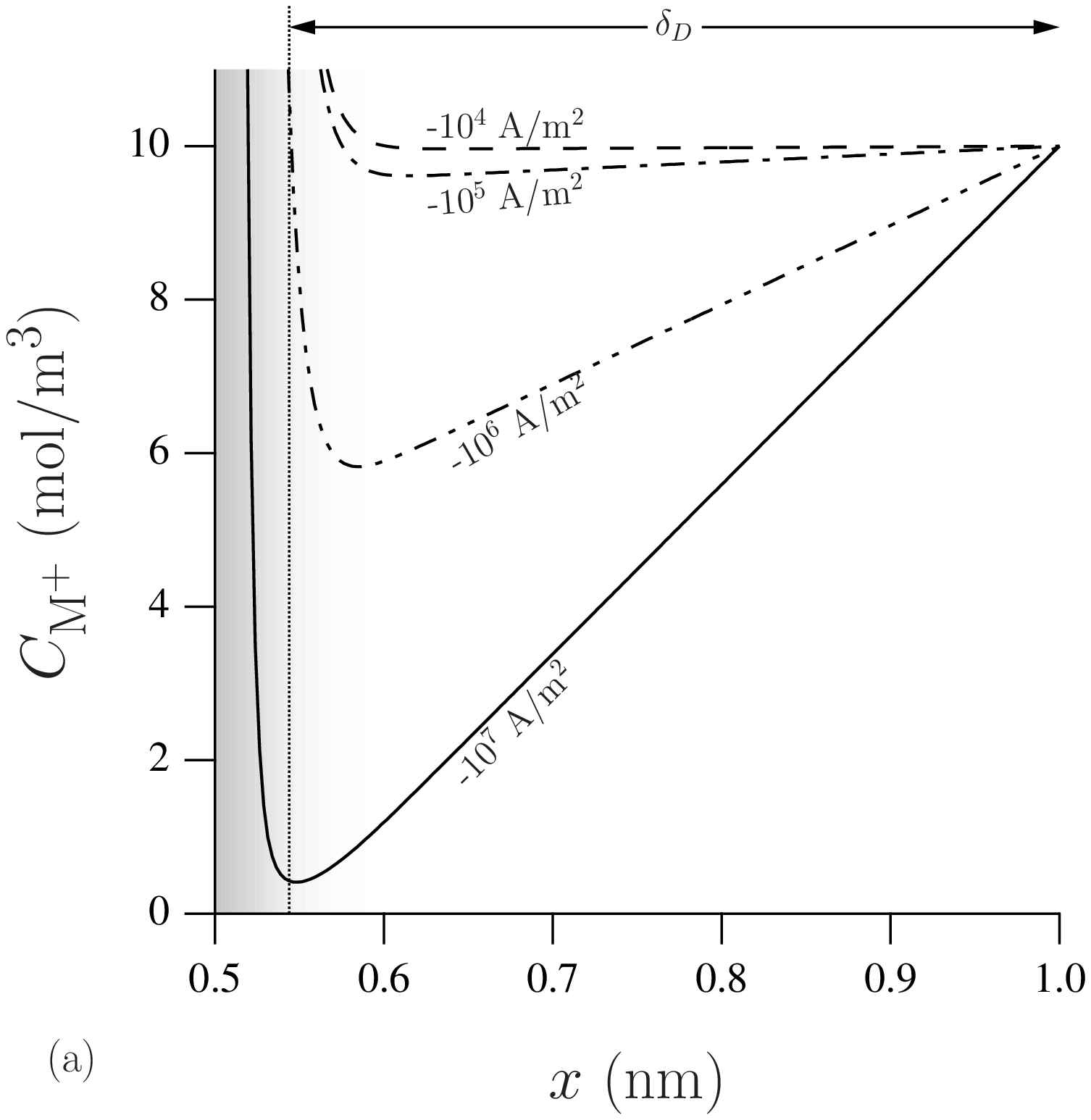}%
	}%
    \subfigure{\label{fig:concOverpotential:Corroding}%
	\includegraphics[width=\subfigurewidth]{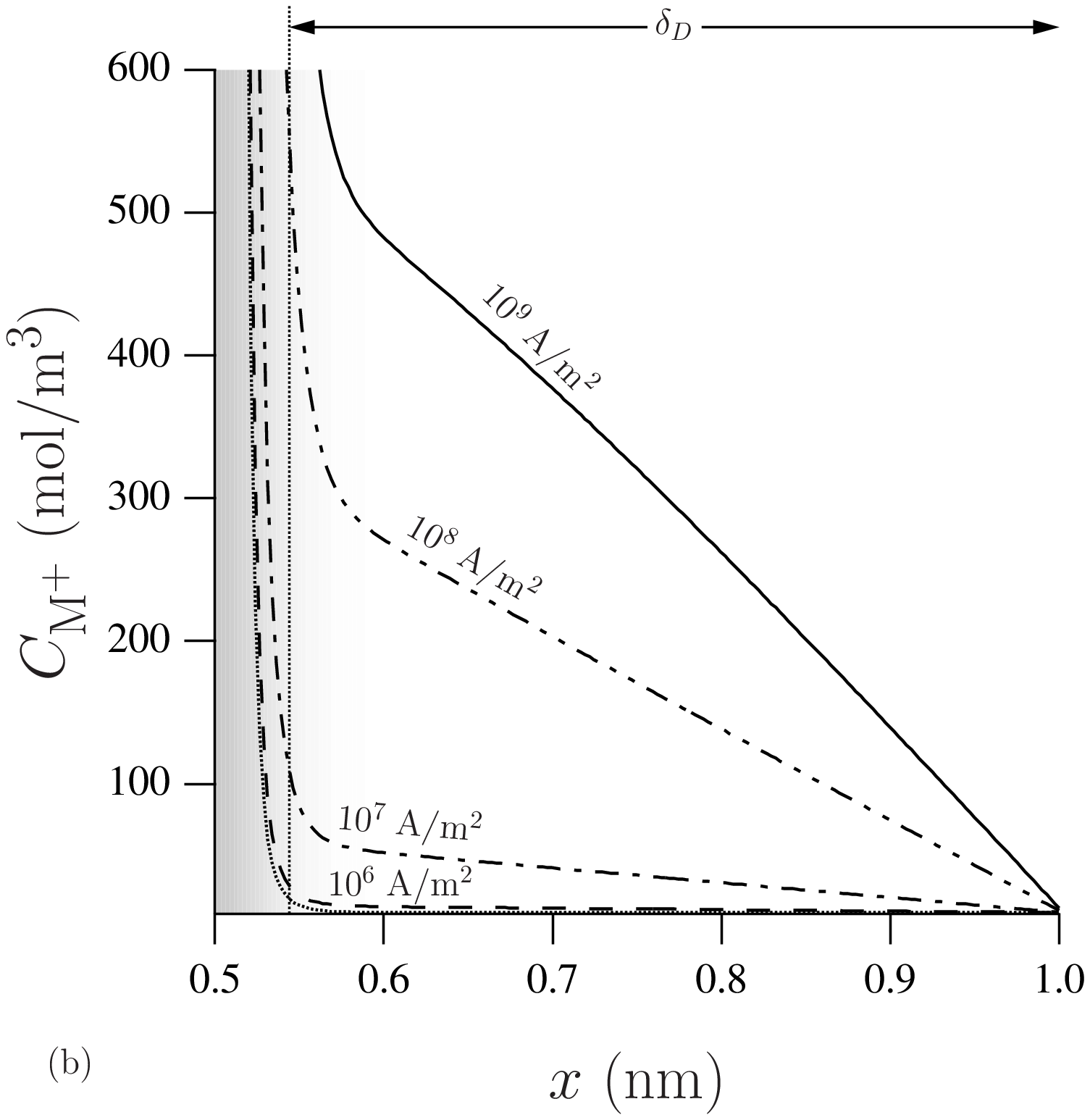}%
	}%
    \caption{Concentration of \Cation{+} as a function of position in the
    electrolyte for different total \PRE{(a)} electrodeposition and \PRE{(b)}
    electrodissolution currents.  The concentration at the interface exhibits a
    Nernstian shift in concentration with overpotential as the current is
    changed.  The diffusion boundary layer is clearly linear over the small
    simulation domain.  \( \DoubleWell(\Phase) \) is mapped onto the background
    in gray to indicate the location of the \PRE{phase field} interface.  The dashed vertical line
    indicates the thickness of the diffusion boundary layer \(
    \Thickness_{\Diffusivity{}} \) calculated in
    Section~\ref{sec:Parameters:Interface}.  The concentration gradient at \(
    \Current = \unit{\power{-10}{7}}{\amperepersquaremetre} \) gives rise to the
    \Cation{+} limiting current of \( \Current_\text{lim} \approx
    \unit{\Scientific{-2}{6}}{\amperepersquaremetre} \); the majority of the
    current is carried by other species.}
    \label{fig:concOverpotential}
\end{figure*}
In Figure~\ref{fig:concOverpotential} we plot the profile of \Cation{+} in the
electrolyte, showing the depletion due to electrodeposition and the enrichment due to
electrodissolution.  At the highest current in Figure~\ref{fig:concOverpotential:Plating},
we can see that \Concentration{\Cation{+}} near the surface of the electrode is
depleted practically to zero, giving rise to the limiting current behavior of
Section~\ref{sec:Sharp}.  The diffusion layer thickness \(
\Thickness_{\Diffusivity{}} = 0.456\BoxSize \), calculated in
Section~\ref{sec:Parameters:Interface}, is indicated for comparison.  Over the
range of applied currents examined, the enrichment of \Cation{+} during electrodissolution
is not similarly constrained.

\subsection{Current-Overpotential Relationship}

\begin{figure*}[tbp]
    \centering
    \subfigure{\label{fig:CurrentOverpotential:Linear}%
	\includegraphics[width=\subfigurewidth]{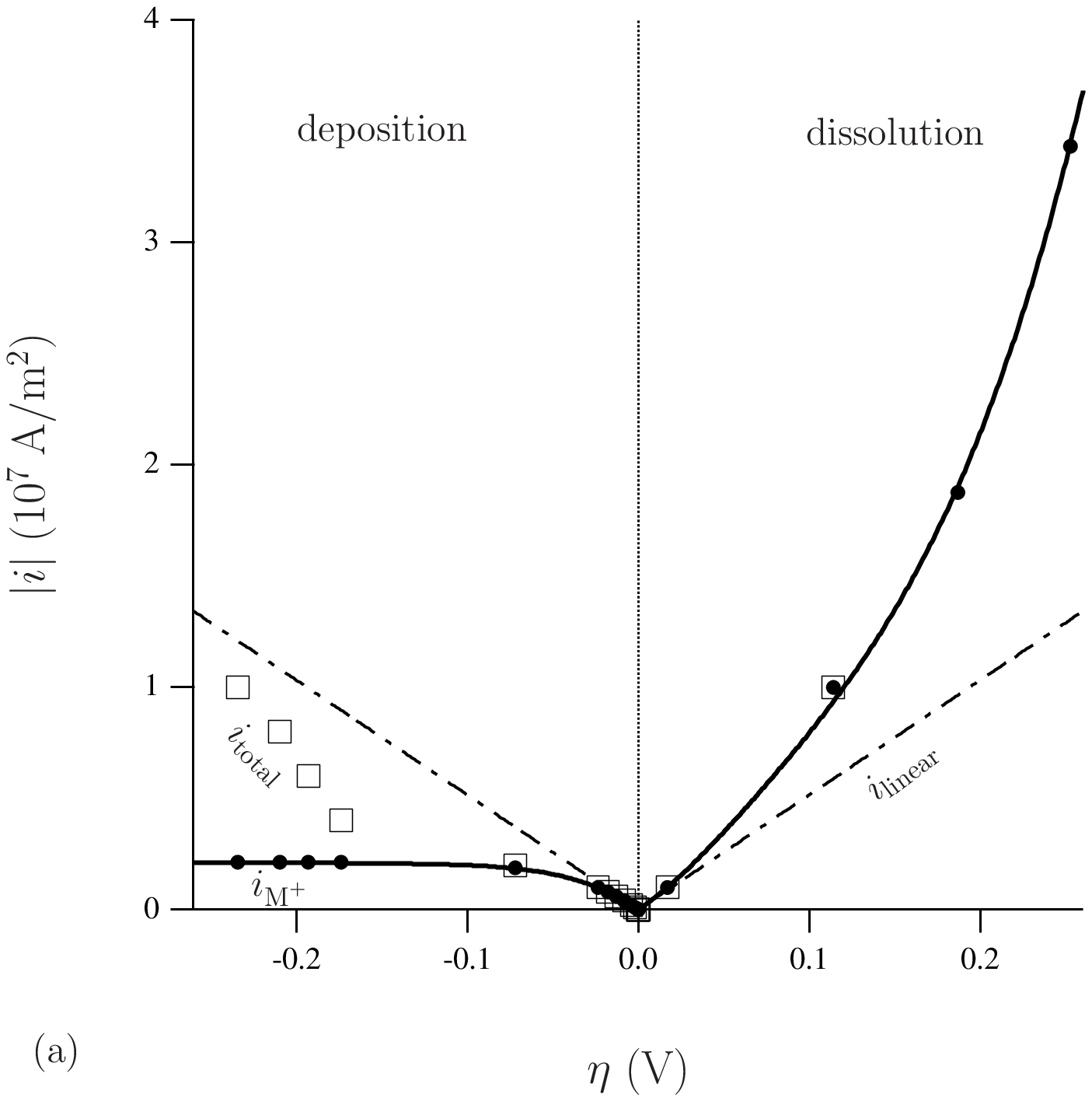}%
	}%
    \subfigure{\label{fig:CurrentOverpotential:Log}%
	\includegraphics[width=\subfigurewidth]{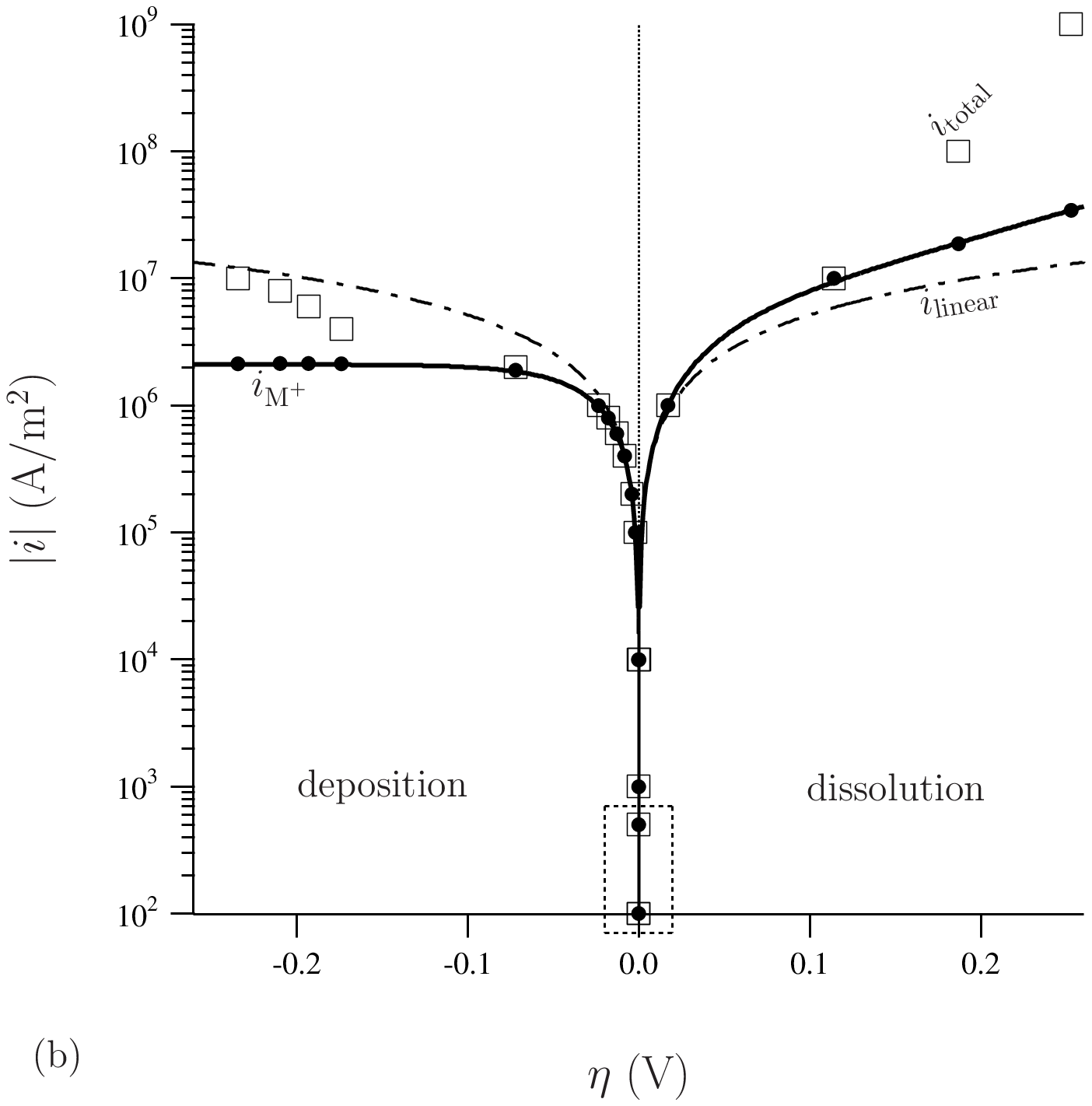}%
	}%
    \caption{Magnitude of the current \Current\ as a function of the
    overpotential \Overpotential, \PRE{plotted on (a) linear and (b) log-linear
    scales}.  The open squares indicate the total current and the circles
    indicate the partial current of \Cation{+}.  The solid line is a plot of the
    current-overpotential equation \eqref{eq:CurrentOverpotential} for \(
    \Current_{0} = \unit{\Scientific{3.80}{6}}{\amperepersquaremetre} \), \(
    \Current_\text{lim} = \unit{\Scientific{-2.15}{6}}{\amperepersquaremetre} \),
    and \( \Transfer = 0.777 \).  The dashed line is a plot of the linear
    current-overpotential equation \eqref{eq:CurrentOverpotential:Linear} for the
    same parameters.  The dotted box indicates the points that contribute to
    Figure~\ref{fig:LinearKinetics}.}
    \protect\label{fig:CurrentOverpotential}
\end{figure*}
In Section~\ref{sec:Parameters:Interface}, we found that the relationship between
current \Current\ and overpotential \Overpotential\ in our calculations is
satisfied by the linear relationship \eqref{eq:CurrentOverpotential:Linear} when
\Current\ and \Overpotential\ are small.  Now we plot \Current\ vs.
\Overpotential\ over a larger range of applied currents in
Figure~\ref{fig:CurrentOverpotential} as open squares.
Equation~\eqref{eq:CurrentOverpotential} considers only the electroactive
species, so the filled circles in Figure~\ref{fig:CurrentOverpotential} show the
current carried by the electroactive cation \( \Current_{\Cation{+}} \).  The
relationship between \( \Current_{\Cation{+}} \) and \Overpotential\ is not
linear.  At large, negative values of \Overpotential, we observe a limiting
current, whereas for large positive values of \Overpotential, no such limiting
current is observed and \( \Current_{\Cation{+}} \) appears exponentially
dependent on \Overpotential.  We fit Eq.~\eqref{eq:CurrentOverpotential} to the
calculated values of \( \Current_{\Cation{+}} \) and we find that \( \Current_{0}
= \unit{\Scientific{(3.80\pm 0.08)}{6}}{\amperepersquaremetre} \), \(
\Current_\text{lim} = \unit{\Scientific{(-2.15\pm
0.06)}{6}}{\amperepersquaremetre} \), and \( \Transfer = 0.777\pm 0.002 \).
These values of \( \Current_{0} \) and \( \Current_\text{lim} \) are within 5\%
of the values found in the linear analysis of
Section~\ref{sec:Parameters:Interface}.  Because \( \Current_{0} \) is of the
same order as \( \Current_{\text{lim}} \) in our calculations, we do not observe
an obvious ``Tafel slope'' during electrodeposition.  Nonetheless, the transition
between low current and diffusion-limited current cannot be fit except by the
full form of Eq.~\eqref{eq:CurrentOverpotential}.  From these results, we see
that despite postulating a linear evolution equation for the phase field
(Eq.~\eqref{eq:Evolution:Phase}), we obtain the nonlinear current-overpotential
behavior predicted by sharp-interface theories and observed in electrochemical
experiments.

The transfer coefficient \Transfer\ characterizes the symmetry of the energy
barrier between the electrode and electrolyte phases.  A value of \( \Transfer =
0.5 \) would mean the energy barrier is symmetric and that a given change in
potential would cause the barrier to electrodeposition to change by the same
magnitude as the barrier to electrodissolution.  Our observed value of \(
\Transfer = 0.78 \) indicates that the barrier to electrodeposition is more
sensitive to changes in potential than is the barrier to electrodissolution.
Although we do not know the functional relationship between \Transfer\ and the
parameters of the phase field model, we can surmise that it is related to the
height \Barrier{j} and shape \( \DoubleWell(\Phase) \) of the interfacial energy
barriers.  This will be investigated in the future.

Since the exchange current is equal to the balanced anodic and cathodic current 
passed at equilibrium, it can be shown that \cite{Vetter:1967,Bard:2nd}
\begin{equation}
    \Current_{0} \equiv 
    \RateConstant_{0}\Faraday
    {\Concentration{O}^{\infty}}^{(1-\Transfer)}
    {\Concentration{R}^{\infty}}^{\Transfer}.
    \label{eq:ExchangeCurrent:Definition}
\end{equation}
\( \Concentration{O}^{\infty} \) is the concentration of the oxidized
electroactive species in the bulk electrolyte, which is \(
\Concentration{\Cation{+}}^{\Electrolyte} = \unit{10}{\molepercubicmetre} \) in
our notation.  \( \Concentration{R}^{\infty} \) is the concentration of the
reduced electroactive species in the bulk electrode, which is \(
\Concentration{\Cation{+}}^{\Electrode} \approx
1/\PartialMolarVolume{\Substitutional} \) in our notation.  The only terms we
cannot directly identify in our phase field model are the dimensionless transfer
coefficient \Transfer\ and the rate constant \( \RateConstant_{0} \).  Noting
that we found \( \Current_{0} \propto \Mobility{\Phase} \) in
Section~\ref{sec:Parameters:Interface}, from a dimensional analysis, one may
expect that
\begin{equation}
    \RateConstant_{0} \propto \Mobility{\Phase}\SurfaceEnergy.
    \label{eq:RateConstant:Proportionality}
\end{equation}
The surface \PRE{free} energy found in our paper on the equilibrium electrochemical
interface \cite{ElPhF:equilibrium} is
\begin{equation}
    \SurfaceEnergy
    = \int_{-\infty}^{\infty} 
    \left[
	\Gradient{\Phase}\left(\Phase'\right)^2
	- \Dielectric \left(\Potential'\right)^2
    \right] \, d x.
    \label{eq:SurfaceEnergy}
\end{equation}
From numerical calculations on the system in this paper when \( \Current = 0 \),
we obtain a value of \( \SurfaceEnergy = \unit{0.46}{\joulepersquaremetre} \).
If we assume that \( \RateConstant_{0} \) in
Eq.~\eqref{eq:RateConstant:Proportionality} is not just proportional to but equal
to \( \Mobility{\Phase}\SurfaceEnergy \) and substitute this value of the
surface \PRE{free} energy and Eq.~\eqref{eq:ExchangeCurrent:Fit} into
Eq.~\eqref{eq:ExchangeCurrent:Definition}, we obtain \( \Transfer \approx 0.73
\).  If we assume instead that the surface \PRE{free} energy is that found in models of
single component solidification \cite{Wheeler:1992}
\begin{align}
    \SurfaceEnergy 
    &= \sqrt{\frac{\Gradient{\Phase}\Barrier{}}
	{18\PartialMolarVolume{\Substitutional}}} = \unit{0.38}{\joulepersquaremetre},
    \label{eq:Phase:SurfaceEnergy}
\end{align}
we find that \( \Transfer \approx 0.75 \).  In either case, this value of
\Transfer\ is very close to that obtained by comparing our results to the the
sharp interface equation \eqref{eq:CurrentOverpotential}, and is not strongly
sensitive to the choice of \SurfaceEnergy. Although \Transfer\ is usually 
assumed to be \( 1/2 \) when no other information is available, it can take on 
any value between 0 and 1 for an ion transfer reaction \cite{Schmickler:1996}.

\subsection{Alloy Electrodeposition}


\ifx\undefined\pdfoutput
\begin{figure*}[p]
    \centering
    \includegraphics[height=0.9\textheight]{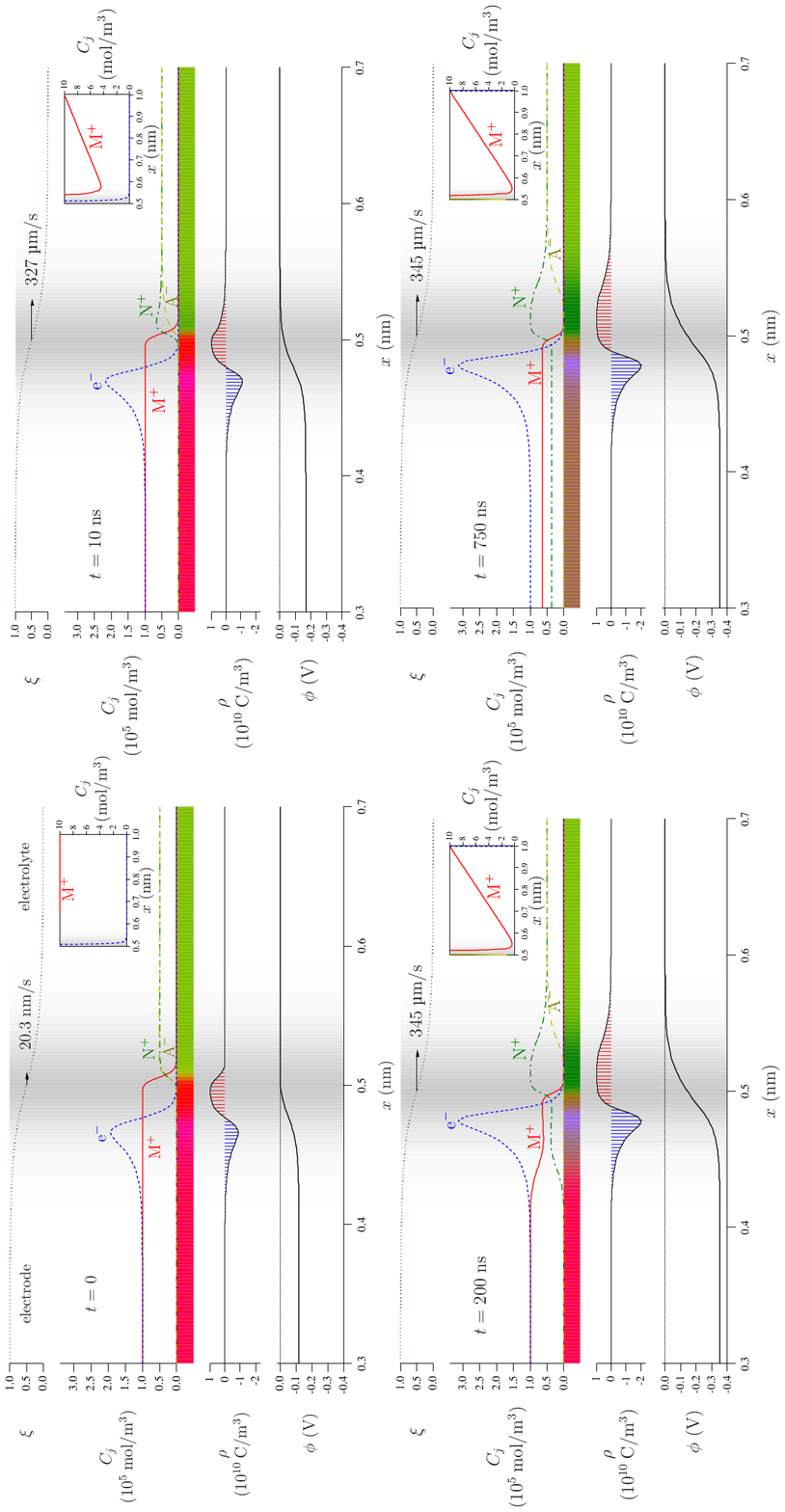}
    \caption{Progress of alloy electrodeposition upon step from \( \Current =
    \unit{\power{-10}{2}}{\amperepersquaremetre} \) to \( \Current =
    \unit{\power{-10}{7}}{\amperepersquaremetre} \).  \PRE{\(
    \DoubleWell(\Phase) \) is mapped onto the background in gray to indicate the
    location of the phase field interface.}  \PRE{(Color)}}
    \label{fig:alloy:sequence}
\end{figure*}
\else
\begin{turnpage}
    \begin{figure*}[p]
	\centering
	\includegraphics[width=\textheight]{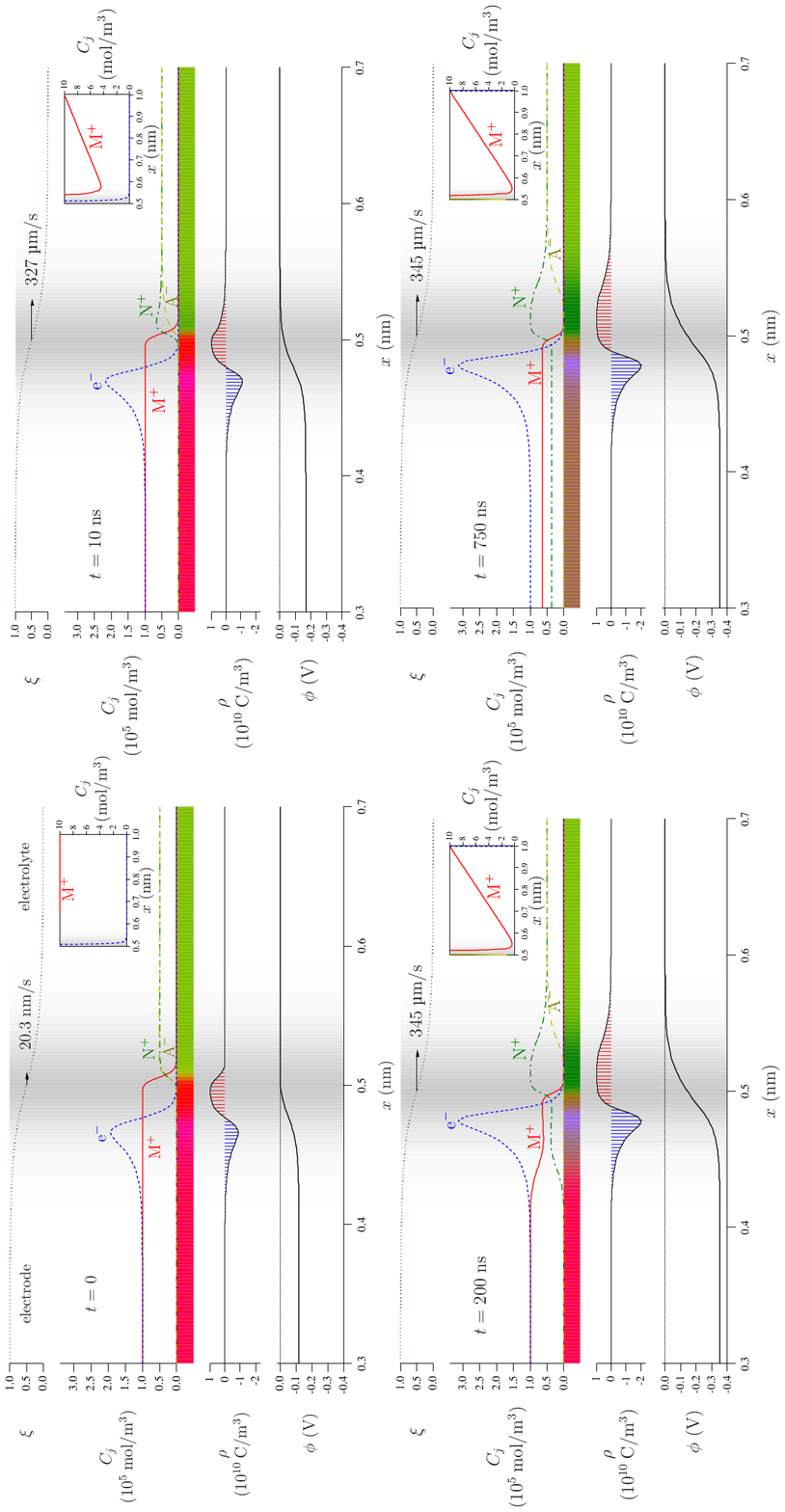}
	\caption{Progress of alloy electrodeposition upon step from \( \Current =
	\unit{\power{-10}{2}}{\amperepersquaremetre} \) to \( \Current =
	\unit{\power{-10}{7}}{\amperepersquaremetre} \).  \PRE{\(
	\DoubleWell(\Phase) \) is mapped onto the background in gray to indicate
	the location of the phase field interface.} \PRE{(Color)}}
	\label{fig:alloy:sequence}
    \end{figure*}
\end{turnpage}
\fi

We examine electrodeposition of alloys by increasing the applied current by five
orders of magnitude from \unit{\power{-10}{2}}{\amperepersquaremetre} to
\unit{\power{-10}{7}}{\amperepersquaremetre}, starting from the steady state
result of Figure~\ref{fig:SteadyState}.  The fields in the vicinity of the
interface are displayed at four different times in
Figure~\ref{fig:alloy:sequence}.  We have added a small concentration inset to
each frame to highlight the behavior of \Cation{+} in the electrolyte and a bar
of color that represents the overall composition of the system.  The initial
potential drop across the interface of \( \Galvani{} = \unit{0.118}{\volt} \) is
within \unit{2}{\micro\volt} of the Nernst potential for \(
\Concentration{\Cation{+}}^{\Electrolyte} = \unit{10}{\mole\per\cubic\meter} \).
At \unit{10}{\nano\second} after the step in current, \Concentration{\Cation{+}}
has depleted at the interface to approximately half its bulk value and
\Otherion{+} has begun to accumulate at the electrode surface.  At
\unit{200}{\nano\second}, \Concentration{\Cation{+}} has depleted essentially to
zero at the electrode surface, giving rise to the limiting current of \Cation{+}
through the electrolyte.  This \Cation{+} current of approximately
\unit{\Scientific{-2.1}{6}}{\amperepersquaremetre} is not adequate to meet the
applied current of \unit{\power{-10}{7}}{\amperepersquaremetre}.  The surface of
the electrode becomes covered with a layer very rich in \Otherion{+} and an alloy
of \Cation{} and \Otherion{} begins to deposit on the electrode.  By
\unit{750}{\nano\second}, the interfacial structure established at
\unit{200}{\nano\second} is essentially unchanged and the original, pure
\Cation{} electrode has been completely swept from view, replaced by a
\Cation{}\Otherion{} alloy.

\begin{figure}[tbp]
    \centering
    \includegraphics[width=\figurewidth]{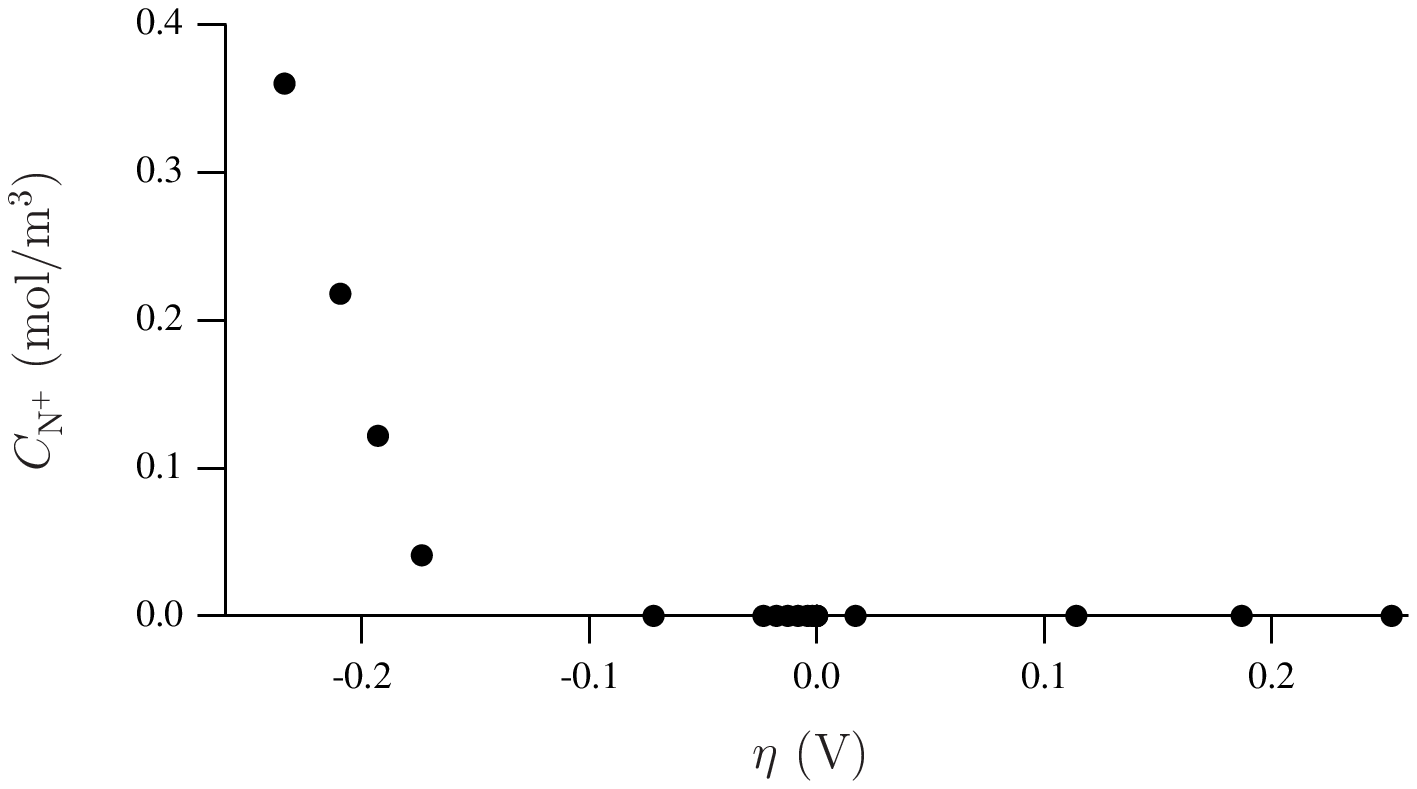}
    \caption{Concentration \Concentration{\Otherion{+}} of \Otherion{+} in the
    deposited electrode as a function of overpotential \Overpotential.}
    \protect\label{fig:alloy:concentration}
\end{figure}
In Figure~\ref{fig:alloy:concentration}, we plot the steady-state concentration
of \Otherion{+} in the electrode as a function of \Overpotential. For small 
overpotentials, up to \( \Overpotential \approx \unit{-0.17}{\volt} \), the 
electrode is essentially pure \Cation{}. At large magnitudes of \Overpotential, 
the fraction of \Otherion{} grows in an apparently linear fashion.

\subsection{Interface Structure}

The concentration and charge distributions at the interface are sensitive to the
electrodeposition conditions at all overpotentials or applied currents, but can be seen
clearly in Figure~\ref{fig:alloy:sequence}.  At
\( \Current = \unit{\power{-10}{2}}{\amperepersquaremetre} \), \Concentration{\Cation{+}},
\Concentration{\Otherion{+}} and \Concentration{\Anion{-}} in the electrolyte
remain very close to their bulk values, all the way into the interfacial region.
The charge distribution consists of a dipole on the electrode side, with very
small net negative charge, and a corresponding positive charge on the
electrolyte.  At \( \Current = \unit{\power{-10}{7}}{\amperepersquaremetre} \),
\Concentration{\Cation{+}} is depleted nearly to zero at the interface and
\Otherion{+} displaces essentially all of the \Anion{-} at the interface.  The
density of \Electron\ at the surface of the electrode is much larger than at the
lower current and the charge distribution has shifted to a predominantly negative
charge on the electrode and a positive charge on the electrolyte.  These changes
in the charge distribution are directly tied to the change in overpotential,
through Eq.~\eqref{eq:Governing:Poisson}.

\section{Conclusions}

Previously \cite{ElPhF:equilibrium}, we developed a phase field model of the
electrochemical interface.  We performed numerical calculations on a model system
like an aqueous electrolyte, in which the majority species in the electrolyte had
no charge.  We demonstrated that, even with a simple ideal solution thermodynamic
description, our model exhibited charged double layer behavior, an
``electrocapillary'' relationship between surface \PRE{free} energy and electrostatic
potential difference across the interface, and differential capacitance curves
that are strongly reminiscent of experimental measurements.

In this paper, we have applied the same phase field model to electrodeposition
and electrodissolution conditions.  We have performed numerical calculations on a
model system like a molten salt, with four species which all carry charge.  We
have shown:
\begin{itemize} 
    
    \item the relationship between the parameters of the phase
    field model and the physical parameters of an electrochemical system,

    \item that our model electrode carries current by electromigration of 
    electrons and that our model electrolyte carries current by diffusion of 
    cations,

    \item that the diffusion field in the electrolyte is essentially linear and 
    that limiting current behavior results,

    \item that despite making linear postulates for the time-dependent governing
    equations, the current-overpotential relationship is
    non-linear and agrees very well with the classic sharp-interface 
    relationship (``Butler-Volmer'' with mass transport effects),

    \item that currents in excess of the limiting current for the more noble
    species result in the deposition of alloys,
    
    \item that there are changes in the double layer structure with current.
\end{itemize}

As discussed in Ref.~\cite{ElPhF:equilibrium}, the need to resolve the charge
distribution in close proximity to the interface limits the size of the domain
and the time spans we can model.  Possibly, adaptive mesh techniques and implicit
solution methods will enable us to examine larger domains and longer times.
Nonetheless, our work here demonstrates that the phase field approach, using a
very simple set of assumptions, can reproduce the rich behaviors of existing
electrochemical theories and permit exploration of the relationship between
double layer structure and interfacial kinetics.

\begin{acknowledgments}
    The authors are grateful for patient explanations of electrochemistry by
    U.~Bertocci, E.~Gileadi, T.~P.~Moffat, and G.~R.~Stafford, for
    helpful discussions regarding the modeling of phase transformations with
    J.~W.~Cahn, S.~Coriell, A.~Lobkovsky, R.~F.~Sekerka, and D.~Wheeler, and for
    a careful reading of this manuscript by E.~Garc\'\i a.  Part of this
    research was supported by the Microgravity Research Division of NASA.
\end{acknowledgments}

\bibliography{abbrTitles,electrochemistry,phaseField,diffusion}
\bibliographystyle{apsrev}

\end{document}